\documentclass{jfm}

\usepackage{graphicx}
\usepackage{newtxtext}
\usepackage{newtxmath}
\usepackage{natbib}
\usepackage{hyperref}
\usepackage{multirow}
\hypersetup{
    colorlinks = true,
    urlcolor   = blue,
    citecolor  = blue,
}

\newcommand{\RomanNumeralCaps}[1]
\nolinenumbers

\makeatletter

\providecommand{\tabularnewline}{\\}

\title[3D Buoyant HF: finite volume]{Three-dimensional buoyant hydraulic fractures: finite volume release}
\author[A. M\"ori, B. Lecampion]{Andreas M\"ori and Brice Lecampion\thanks{Email for correspondence: brice.lecampion@epfl.ch, andreas.mori@epfl.ch} } 
\affiliation{Geo-Energy Laboratory - Gaznat Chair on Geo-Energy,\\  
Ecole Polytechnique F\'ed\'erale de Lausanne, \\
 ENAC-IIC-GEL-EPFL,  Station 18, CH-1015, Switzerland}

\pubyear{2023} 
\volume{--} 
\pagerange{--} 
\date{received --, revised --, accepted --}  

\makeatother

\begin{document}
\title{Three-dimensional buoyant hydraulic fractures: finite volume release}
\maketitle
\begin{abstract} 
In impermeable media, a hydraulic fracture can continue to expand even without additional fluid injection if its volume exceeds the limiting volume of a hydrostatically loaded radial fracture. This limit depends on the mechanical properties of the surrounding solid and the density contrast between the fluid and the solid. Self-sustained  fracture growth is characterized by two dimensionless numbers. The first parameter is a buoyancy factor that compares the total released volume to the limiting volume to determine whether buoyant growth occurs. The second parameter is the dimensionless viscosity of a radial fracture at the time when buoyant effects become of order 1. This dimensionless viscosity notably depends on the rate at which the fluid volume is released, indicating that both the total volume and release history impact self-sustained buoyant growth. Six well-defined propagation histories can be identified based on these two dimensionless numbers.  Their growth evolves between distinct limiting regimes of radial and buoyant propagation, resulting in different fracture shapes. We can identify two growth rates depending on the dominant energy dissipation mechanism (viscous flow vs fracture creation) in the fracture head. For finite values of material toughness, the toughness-dominated limit represents a late-time solution for all fractures in growth rate and head shape (possibly reached only at a very late time). The viscosity-dominated limit can appear at intermediate times. Our three-dimensional simulations confirm the predicted scalings and highlight the importance of considering the entire propagation and release history for accurate analysis of buoyant hydraulic fractures.

\end{abstract}

\section{Introduction}

This work investigates the growth of a planar three-dimensional (3-D)
hydraulic fracture (HF) from the release of a finite volume of fluid
from a point source and its possible transition to a self-sustained
buoyant fracture. Hydraulic fractures are tensile, fluid-filled fractures driven by the internal
fluid pressure exceeding the minimum compressive in-situ stress \citep{Deto16}.
Natural occurrences of HFs are related to the transport of magma through
the lithosphere by magmatic intrusions \citep{RiTa15,SpSh87,LiKe91}
or pore pressure increases due to geochemical reactions during the
formation of hydrocarbon reservoirs \citep{Vern94}. One of the most
frequent engineering applications of HFs relates to the production stimulation of hydrocarbon wells \citep{EcNo00,SmMo15,JeCh13}.

In the absence of buoyancy, the propagation of radial hydraulic fractures
upon the end of the release (denoted as "shut-in'' in industrial
applications) has recently been analysed in detail \citep{MoLe21}.
In an impermeable media, the final radius of the HF solely depends
on the material parameters and the total amount of fluid volume injected/released.
However, the HF does not necessarily stop its growth directly upon
the end of the release. When dissipation through viscous fluid flow
is important at the end of the release, the propagation continues
for a while in a viscosity-dominated pulse regime before finally arresting
at an arrest radius independent of the release rate.

\begin{figure}
\centering{}\includegraphics[width=0.95\textwidth]{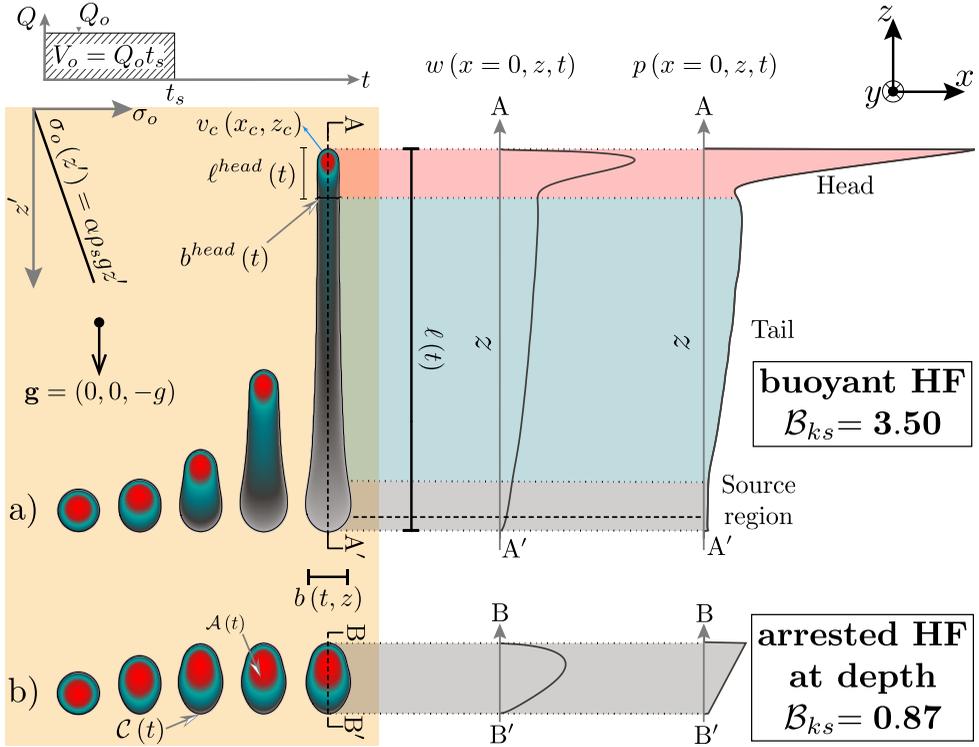}\caption{a)
Buoyant self-sustained growth of a hydraulic fracture.
b) Arrested hydraulic fracture at depth. Both fractures emerge from
a finite fluid volume released from a point source through a block
injection and propagate in a homogeneous
linear elastic medium ($x\vert z$ plane) with the downwards oriented
gravity vector ${\displaystyle \mathbf{g}}$ (in ${\displaystyle -z}$)
creating a linear confining stress ${\displaystyle \sigma_{o}\left(z\right)}$.
Fracture area is denoted by ${\displaystyle \mathcal{A}\left(t\right)}$
with a closed front ${\displaystyle \mathcal{C}\left(t\right)}$ and
a local normal velocity ${\displaystyle v_{c}\left(x_{c},z_{c}\right)}$
(with ${\displaystyle \left({\displaystyle x_{c},}z_{c}\right)\in{\displaystyle \mathcal{C}\left(t\right)}}$).
The fracture extent is defined by its local breadth ${\displaystyle b\left(z,t\right)}$
and total length ${\displaystyle \ell\left(t\right)}$. \label{fig:Sketch}}
\end{figure}

When considering gravity, recent research has focussed on the derivation
of the limiting volume necessary for the emergence of a three-dimensional buoyant fracture
\citep{Dahm00,DaRi20,SaZi20,SmPi21}. Neglecting fluid viscosity, \citet{DaRi20} identify
a critical volume similar to previous two-dimensional (2-D) predictions \citep{Weer71}. It
is, however, not possible to constrain the ascent rate of the fracture
without accounting for the effect of fluid viscosity (as discussed
in \citet{GaGe14}). The consensus of these studies is that the resulting buoyant fracture
features a head and tail structure \citep{LiKe91}, where the head
dominates the overall fracture behaviour, but the tail dominates the
ascent rate \citep{GaGe22} (see figure \ref{fig:Sketch}). \citet{DaRi23}
estimate a maximum ascent velocity considering a viscosity-dominated
tail. A similar solution has been derived by \citet{GaGe14} (see
\citet{GaGe22} for details) for a finger-like fracture with a toughness-dominated
head. In their work, they derive a three-dimensional (3-D) head similar
to the limiting volume of \citet{DaRi20}. This fracture "head"
is coupled to a tail of constant breath, providing a late-time solution
after the end of the transition from radial to self-sustained buoyant
propagation. Considering lubrication flow in the initially radially
propagating fracture, \citet{SaZi20} performed a few simulations
investigating the early phase of the transition to buoyant propagation. Equivalent
to \citet{DaRi20} and \citet{GaGe14}, a limiting value for the necessary
volume released for a buoyant fracture to emerge is reported. All
three estimates of the minimal/critical volume release have the same characteristic
scale and only differ in pre-factors. A combined study of the limiting
volume, considering not only the emergence of buoyancy-driven fractures
but also their evolution towards their late-time characteristics,
is not yet available. 

\section{Preliminaries\label{sec:Preliminaries}}

We investigate tensile (mode I) hydraulic fractures under the classical
assumption of linear elastic fracture mechanics (LEFM) and laminar
Newtonian lubrication flow \citep{Deto16}. The finite volume is released
from a point source at depth into a linearly elastic and impermeable
medium with uniform properties. The fracture orientation and stress
state are equivalent to the one described in \citet{MoLe22c} and
sketched in figure \ref{fig:Sketch}. We omit the detailed discussion
of the mathematical formulation (see \citet{MoLe22c} for details)
as the only difference pertains to the history of the fluid release.We
consider here a simple injection history where the fluid volume is
released at a constant rate until the end of the release at time $t=t_{s}$
(the shut-in time), where the rate suddenly drops to zero. We denote
the constant release rate during the block injection as $Q_{o}$ such
that the rate history is simply:
\begin{equation}
Q\left(t\right)=\begin{cases}
Q_{o} & t\leq t_{s}\\
0 & t>t_{s}
\end{cases}.\label{eq:Qoft}
\end{equation}
The coherent global volume balance in the case of an impermeable medium
is
\begin{equation}
\mathcal{V}\left(t\right)=\int_{\mathcal{A}(t)}w\left(t,x,z\right)\text{d}x\text{d}z=\begin{cases}
Q_{o}t & t<t_{s}\\
V_{o}=Q_{o}t_{s} & t\geq t_{s}
\end{cases},\label{eq:VolumeBalance}
\end{equation}
where $V_{o}=Q_{o}t_{s}$ is the total volume of fluid released.

In the following, we combine scaling arguments and numerical simulations
using the fully-coupled planar 3-D hydraulic fracture solver PyFrac
\citep{ZiLe20}. We refer the reader to \citet{ZiLe20} and references
therein for a detailed description of the numerical scheme. The documentation
of the open-source code and examples of applications are available
for download at \href{https://github.com/GeoEnergyLab-EPFL/PyFrac/releases/tag/JFM2023}{PyFrac}.
The initial conditions and parameters used for the numerical simulations
are detailed in section 2.2 of \citet{MoLe22c} as well as in the
shared data of this article.

\subsection{Arrest of a finite volume radial hydraulic fracture without buoyancy\label{subsec:FVRadial}}

In the absence of buoyant forces, considering the limiting case of
an impermeable medium, hydraulic fractures finally arrest after the
end of the injection when reaching an equilibrium between the injected
volume and the linear elastic fracture mechanics propagation condition.
This problem was investigated in \citet{MoLe21}. The fracture characteristics
at arrest are independent of the shut-in time $t_{s}$. They only
depend on the properties of the solid and the total amount of fluid
released. For example, the arrest radius $R_{a}$ (subscript $a$
for arrest) is given by
\begin{equation}
{\displaystyle R_{a}=\left(\frac{3}{8\sqrt{\pi}}\right)^{2/5}\left(\frac{E^{\prime}V_{o}}{K_{Ic}}\right)^{2/5}},\label{eq:RadialArrestRadius}
\end{equation}
where $E^{\prime}=E/\left(1-\nu^{2}\right)$ is the plane-strain modulus
with $E$ the materials Young's modulus and $\nu$ its Poisson's ratio,
and ${\displaystyle K_{Ic}}$ the fracture toughness of the material.

Even though the arrest radius is independent of $t_{s}$, the growth
history prior to arrest depends on it. In particular, the arrest is
not necessarily immediate after the end of the release. This is notably
the case when the hydraulic fracture propagates in the viscosity-dominated
regime at the end of the release. The immediate arrest versus continuous
growth is captured by the value of the dimensionless toughness at
the shut-in time:
\begin{equation}
\mathcal{K}_{ms}=K_{Ic}\frac{t_{s}^{1/9}}{E^{\prime13/18}\mu^{\prime5/18}Q_{o}^{1/6}}.\label{eq:ShutInToughness}
\end{equation}
where $\mu^{\prime}=12\mu$ and $\mu$ is the fracturing fluid viscosity.
In (\ref{eq:ShutInToughness}), we have used the subscripts $m$ and
$s$ to indicate respectively a viscous scaling and the end of the
release. If the fracture is viscosity-dominated ( $\mathcal{K}_{ms}\ll1$),
it propagates in a viscosity-dominated pulse regime for a while until
it finally arrests when reaching $R=R_{a}$. On the other hand, if
fracture energy is already dominating ( $\mathcal{K}_{ms}\gg1$),
the arrest is immediate upon shut-in. The viscosity-dominated pulse
regime has been shown to emerge for $\mathcal{K}_{ms}\lessapprox0.3$
(for a detailed description of the viscosity-dominated pulse regime,
see section 3.2 of \citet{MoLe21}). A numerical estimation of the
immediate arrest yields a value of $\mathcal{K}_{ms}\gtrapprox0.8$
(note that \citet{MoLe21} report a value of $2.5$ due to an alternative
definition of (\ref{eq:ShutInToughness}) using $K^{\prime}=\sqrt{32/\pi}K_{Ic}$
instead of $K_{Ic}$).

\subsection{Buoyant hydraulic fracture under a continuous release\label{subsec:BuoyantHFCont}}

In the case of a fluid release occurring at a constant volumetric
rate $Q_{o}$, the fracture elongates in the direction of gravity
in the presence of buoyant forces. Two dimensionless buoyancies related
either to the viscosity dominated (subscript $m$) or the toughness-dominated
regime (subscript $k$) emerge \citep{MoLe22c}:

\begin{equation}
\mathcal{B}_{m}=\varDelta\gamma\frac{Q_{o}^{1/3}t^{7/9}}{E^{\prime5/9}\mu^{\prime4/9}},\quad\mathcal{B}_{k}=\varDelta\gamma\frac{E^{\prime3/5}Q_{o}^{3/5}t^{3/5}}{K_{Ic}^{8/5}}.\label{eq:DimlessBuoyancies}
\end{equation}
These dimensionless buoyancies are related through the dimensionless
viscosity of a radial fracture when buoyancy becomes of order $\mathcal{O}(1)$:
\begin{equation}
\mathcal{M}_{\widehat{k}}=\mu^{\prime}\frac{Q_{o}E^{\prime3}\varDelta\gamma^{2/3}}{K_{Ic}^{14/3}}\label{eq:MkhatDefinition}
\end{equation}
as
\begin{equation}
\mathcal{B}_{k}=\mathcal{B}_{m}^{27/35}\mathcal{M}_{\widehat{k}}^{12/35}.\label{eq:BmFOfBkMkhat}
\end{equation}
Similar to the dimensionless toughness at the end of the release $\mathcal{K}_{ms}$
(\ref{eq:ShutInToughness}), $\mathcal{M}_{\widehat{k}}$ defines
if the transition from a radial to an elongated buoyant fracture occurs
in the viscosity- ($\mathcal{M}_{\widehat{k}}\gg1$) or toughness-dominated
($\mathcal{M}_{\widehat{k}}\ll1$) phase of the radial hydraulic fracture
propagation. A family of solutions emerges as a function of this dimensionless
viscosity $\mathcal{M}_{\widehat{k}}$ as discussed in detail in \citet{MoLe22c}.
Notably, a limiting large toughness solution has been obtained in
\citet{GaGe14} (see details in \citet{GaGe22}). This large toughness
limit is observed for $\mathcal{M}_{\widehat{k}}\leq10^{-2}$ \citep{MoLe22c}
and shows a buoyant finger-like fracture with a constant breadth and
a fixed-volume head. These attributes, combined with a constant injection
rate, lead to a linear growth rate of the buoyant fracture. In an
intermediate range of values for $\mathcal{M}_{\widehat{k}}\in\left[10^{-2},10^{2}\right]$,
the fractures exhibit a uniform horizontal breadth and a finger-like
shape. In this range of $\mathcal{M}_{\widehat{k}}$, the prefactors
(for length, width etc.) become dependent on the dimensionless viscosity
$\mathcal{M}_{\widehat{k}}$ (\ref{eq:ShutInToughness}). Particularly,
an increase in fracture breadth and head volume is observed with increasing
values of $\mathcal{M}_{\widehat{k}}$ . Even larger values of $\mathcal{M}_{\widehat{k}}\geq10^{2}$
generate fractures exhibiting a negligible toughness, buoyant solution
at intermediate times, where the growth of the fracture is sub-linear.
The breadth of these fractures increases for a while before reaching
an ultimately constant value in relation to the non-zero value of
fracture toughness. The fracture's growth rate then becomes constant.
Concurrently, the head and tail structure stabilizes. In the strictly
zero-toughness limit, the breadth continuously increases and the fracture
height growth always remains sub-linear as a consequence of global
volume balance.

\subsection{Hydrostatically loaded radial fracture\label{subsec:LinearLoadedFracture}}

The occurrence of the self-sustained buoyant growth of a finite volume
fracture has been investigated by several authors from the point of
view of the static linear elastic equilibrium of a radial fracture
under a linearly varying load \citep{DaRi20,SaZi20,DaRi23}. Under
the hypothesis of zero viscous flow, the net loading opening the fracture
is equal to the hydrostatic fluid pressure minus the linearly varying
background stress $\sigma_{o}\left(z\right)$. The elastic solution
and the evolution of the stress intensity factor at the upper and
lower tip are known analytically for this loading \citep{TaPa00}.
Adopting a LEFM propagation condition, the stress-intensity factor
(SIF) $K_{I}$ at the upper end is set to the material fracture toughness
$K_{Ic}$. On the other hand, the lower tip SIF is set to zero, allowing
the fracture to close and liberate the volume necessary for further
upward propagation. Enforcing the conditions of $K_{I}=K_{Ic}$ at
the upper and $K_{I}=0$ at the lower tip constrains the limiting
volume to

\begin{equation}
V_{\textrm{limit}}\propto\frac{K_{Ic}^{8/3}}{E^{\prime}\varDelta\gamma^{3/5}}=V_{\widehat{k}}^{\textrm{head}}\label{eq:CriticalVolume}
\end{equation}
This minimal volume for buoyant propagation has been independently
identified in recent contributions \citep{DaRi20,SaZi20,DaRi23}.
Interestingly, it corresponds to the volume $V_{\widehat{k}}^{\text{head}}$
of the toughness-dominated head of a buoyant hydraulic fracture in
the case of a constant release \citep{GaGe14,MoLe22c}.

If the volume of fluid released in the radial fracture is slightly
larger than this value, the upper tip would have a stress intensity
$K_{I}>K_{Ic}$, indicating excess energy leading to upward propagation.
Similarly, the lower end would have $K_{I}<0$ and hence pinch. Small
perturbations of the released volume around this minimum would lead
to either an arrest of the fracture (lower volume) or a departure
of a buoyant fracture (larger volume). Note that when the fracture
volume equals this minimal volume, and fluid viscosity is neglected,
the previous derivation fails to predict how the fracture will subsequently
propagate. Only the introduction of fluid viscosity can resolve the
physical limitation of this approach.

In addition, the previous derivation of the minimum volume for the
occurrence of a self-sustained propagation assumes a perfectly radial
shape until the entire fluid volume has been released. This approach
is equivalent to considering buoyant forces only after the emplacement
of the total released fluid has finished. It does not cover the case
where buoyant forces are non-negligible during the release. It is
further unclear if it applies to the case where fracture growth continues
after the end of the release while buoyant forces are still negligible.

\section{Arrest at depth vs self-sustained propagation of buoyant hydraulic
fractures\label{sec:FractureStagnation}}

From the discussion of the arrest radius of a hydraulic fracture in
the absence of buoyancy (see section \ref{subsec:FVRadial}) and the
regimes of buoyant hydraulic fracture growth under a continuous release
(see section \ref{subsec:BuoyantHFCont}), we can anticipate several
scenarios with respect to the emergence of a self-sustained buoyant
finite volume fracture. The transition towards buoyancy-driven growth
can occur during the release of fluid or during the pulse propagation
phase, when the propagation is viscosity-dominated at the end of the
release. We investigate these different propagation histories in relation
to the dimensionless buoyancies and dimensionless buoyant viscosity
introduced in section \ref{sec:Preliminaries} and discuss their relationship
with the critical minimum volume (\ref{eq:CriticalVolume}).

\subsection{Toughness-dominated at the end of the release\label{subsec:ToughTransLimitDeriv}}

We first investigate the case where the fracture is toughness-dominated
at the end of the release. In the absence of buoyancy, a constant
fluid pressure establishes in the penny-shaped fracture which immediately
stops at its arrest radius $R_{a}$ (see equation (\ref{eq:RadialArrestRadius})).
Due to the addition of buoyant effects, a linear pressure gradient
develops and creates the configuration discussed above (see section
\ref{subsec:LinearLoadedFracture}). We anticipate that the total
volume released must exceed $V_{\widehat{k}}^{\textrm{head}}$ (\ref{eq:CriticalVolume})
for a buoyant fracture to emerge. Neglecting the temporal evolution,
the comparison $V_{o}/V_{\widehat{k}}^{\textrm{head}}$ is sufficient
to assess the emergence of buoyant fractures. When considering a radial
growth in time, the dimensionless buoyancy $\mathcal{B}_{k}\left(t\right)$
(\ref{eq:DimlessBuoyancies}) indicates when buoyant forces become
dominant. Estimating $\mathcal{B}_{k}\left(t\right)$ at the end of
the release $t=t_{s}$, we obtain
\begin{equation}
\mathcal{B}_{ks}=\mathcal{B}_{k}\left(t=t_{s}\right)=\varDelta\gamma\frac{E^{\prime3/5}Q_{o}^{3/5}t_{s}^{3/5}}{K_{Ic}^{8/5}}=\varDelta\gamma\frac{E^{\prime3/5}V_{o}^{3/5}}{K_{Ic}^{8/5}}=\left(\frac{V_{o}}{V_{\widehat{k}}^{\textrm{head}}}\right)^{3/5}.\label{eq:ShutInBuoyancyK}
\end{equation}
From this last relation (\ref{eq:ShutInBuoyancyK}), we see that the
condition of a dimensionless buoyancy at the end of the release $\mathcal{B}_{ks}>1$
(under the hypothesis of a radial toughness-dominated fracture) is
strictly equivalent to the condition of a released volume larger than
the minimal volume for buoyant growth (\ref{eq:CriticalVolume}).

\subsection{Viscosity-dominated at the end of the release $\mathcal{K}_{ms}\ll1$\label{subsec:ViscTransLimitDeriv}}

In contrast to toughness-dominated hydraulic fractures, radial viscosity-dominated
fractures at the end of the release will continue to propagate in
a viscous-pulse regime until they reach their arrest radius $R_{a}$
(\ref{eq:RadialArrestRadius}) \citep{MoLe21}. During that post-release
propagation phase, the fracture may become buoyant and continue its
growth as a result. In addition, we need to check if it remains buoyant
when it is already so at the end of the release. This can be done
by estimating the dimensionless buoyancy of a radial viscous fracture
$\mathcal{B}_{m}\left(t\right)$ (\ref{eq:DimlessBuoyancies}) at
the end of the release $t=t_{s}$:

\begin{equation}
\mathcal{B}_{ms}=\mathcal{B}_{m}\left(t=t_{s}\right)=\varDelta\gamma\frac{Q_{o}^{1/3}t_{s}^{7/9}}{E^{\prime5/9}\mu^{\prime4/9}}=\varDelta\gamma\frac{V_{o}^{1/3}t_{s}^{4/9}}{E^{\prime5/9}\mu^{\prime4/9}}.\label{eq:ShutInBuoyancyM}
\end{equation}
A value of $\mathcal{B}_{ms}\geq1$ indicates that the fracture has
already transitioned to buoyant propagation when the release stops
and is already elongated. On the other hand, if ${\displaystyle \mathcal{B}_{ms}<1}$,
buoyancy is not of primary importance at the end of the release, and
the fracture still exhibits a radial shape.

\subsubsection{Dominant buoyancy at the end of the release $\mathcal{B}_{ms}\protect\geq1$}

In the case $\mathcal{B}_{ms}\geq1$, the fracture is already buoyant
at the end of the release. We must check if it remains buoyant or
possibly arrests after the release ends. It is natural to compare
the volume of the viscous head at the end of the release $V_{\widehat{m}}^{\textrm{head}}\left(t=t_{s}\right)$
to the limiting volume (equation (\ref{eq:CriticalVolume})). The
time-dependent volume of a viscous head is given in equation 5.6 of
\citet{MoLe22c} and relates to equation (\ref{eq:ShutInBuoyancyM})
as
\begin{equation}
\mathcal{B}_{ms}=\left(\frac{V_{o}}{V_{\widehat{m}}^{\textrm{head}}\left(t=t_{s}\right)}\right)^{2/3}.\label{eq:ShutInBuoyancyWithVolumes}
\end{equation}
Using the relationship of equation (\ref{eq:BmFOfBkMkhat}), we obtain
the following relation with the minimal limiting volume:
\[
\frac{V_{o}}{V_{\widehat{k}}^{\textrm{head}}}=\left(\frac{V_{o}}{V_{\widehat{m}}^{\textrm{head}}\left(t=t_{s}\right)}\right)^{6/7}\mathcal{M}_{\widehat{k}}^{4/7}.
\]
For a viscosity-dominated fracture, one necessarily has $\mathcal{M}_{\widehat{k}}\geq1$
and to be buoyant at the end of the release, we necessarily have $V_{o}\geq V_{\widehat{m}}^{\textrm{head}}\left(t=t_{s}\right)$
as $\mathcal{B}_{ms}\geq1$. As a result of the previous relations,
we necessarily have $V_{o}\geq V_{\widehat{k}}^{\textrm{head}}$,
respectively $\mathcal{B}_{ks}\ge1$, and the volume released is larger
than the minimum required for a toughness-dominated radial fracture
subjected to a linear pressure gradient to become buoyant. After the
release has ended, the viscous forces diminish in the head which ultimately
becomes toughness-dominated. As a result, after the release, as buoyancy
is of order one, the condition $\mathcal{B}_{ks}\ge1$ is always satisfied
and self-sustained buoyant growth will necessarily continue.

\subsubsection{Viscosity-dominated fracture with negligible buoyant forces at the
end of the release $\mathcal{B}_{ms}<1$}

If buoyancy forces are negligible at the end of the release, and the
propagation is viscosity-dominated (${\displaystyle \mathcal{B}_{ms}<1}$
and $\mathcal{K}_{ms}\ll1$), the finite volume fracture will continue
to grow radially in a viscous-pulse regime for a while before it finally
arrests. In the presence of buoyant forces, it may be possible that
buoyancy takes over as a driving mechanism before the fracture arrests.
To incorporate such a possible growth history into the analysis, we
use a dimensionless buoyancy in such a radial viscous pulse regime:

\begin{equation}
\mathcal{B}_{m}^{\left[V\right]}\left(t\right)=\varDelta\gamma\frac{V_{o}^{1/3}t^{4/9}}{E^{\prime5/9}\mu^{\prime4/9}}=\mathcal{B}_{ms}\left(t/t_{s}\right)^{4/9},\label{eq:BmPulse}
\end{equation}
where the superscript $\left[V\right]$ indicates that the scaling
is related to a finite volume release (replacing $Q_{o}$ by $V_{o}/t$
in the continuous release expression). From \citet{MoLe21}, we know
that the radial viscous pulse fracture stops propagating when it becomes
toughness-dominated. The corresponding time scale for which $\mathcal{\mathcal{K}}_{m}^{\left[V\right]}$
of a finite volume radial hydraulic fracture in the absence of buoyancy
(see equation 10 of \citet{MoLe21}) becomes of order one and the
fracture arrests is given by 
\begin{equation}
t_{mk}^{\left[V\right]}=\frac{E^{\prime13/5}V_{o}^{3/5}\mu^{\prime}}{K_{Ic}^{18/5}}.\label{eq:DefinitionTmkPulse}
\end{equation}
It is thus possible to check if buoyancy is of order one at this characteristic
time of arrest by estimating the value of the dimensionless buoyancy
$\mathcal{B}_{m}^{\left[V\right]}\left(t\right)$ (\ref{eq:BmPulse})
at $t=t_{mk}^{\left[V\right]}$:
\begin{equation}
\mathcal{B}_{m}^{\left[V\right]}\left(t=t_{mk}^{\left[V\right]}\right)=\varDelta\gamma\frac{E^{\prime3/5}V_{o}^{3/5}}{K_{Ic}^{8/5}}=\left(\frac{V_{o}}{V_{\widehat{k}}^{\textrm{head}}}\right)^{3/5}=\mathcal{B}_{ks}.\label{eq:EquivBmt=00003DtmkNBka}
\end{equation}
Interestingly, this evaluation is strictly equivalent to the comparison
of the limiting $V_{\widehat{k}}^{\textrm{head}}$ with the total
released volume $V_{o}$ (see (\ref{eq:ShutInBuoyancyK})). We conclude
that regardless of the propagation history, the comparison of the
released volume with the limiting volume for toughness-dominated buoyant
growth is sufficient to characterize the emergence of a self-sustained
buoyant hydraulic fracture. In what follows, we use the dimensionless
buoyancy of a radial toughness-dominated finite volume hydraulic fracture
$\mathcal{B}_{ks}$ to quantify the emergence of self-sustained growth
($\mathcal{B}_{ks}>1$). Similarly, the volume ratio $V_{o}/V_{\widehat{k}}^{\textrm{head}}=\mathcal{B}_{ks}^{5/3}$
could also be used.

\subsection{Structure of the solution for a finite volume release\label{subsec:ParametricSpace}}

\begin{figure}
\centering{}\includegraphics[width=0.95\textwidth]{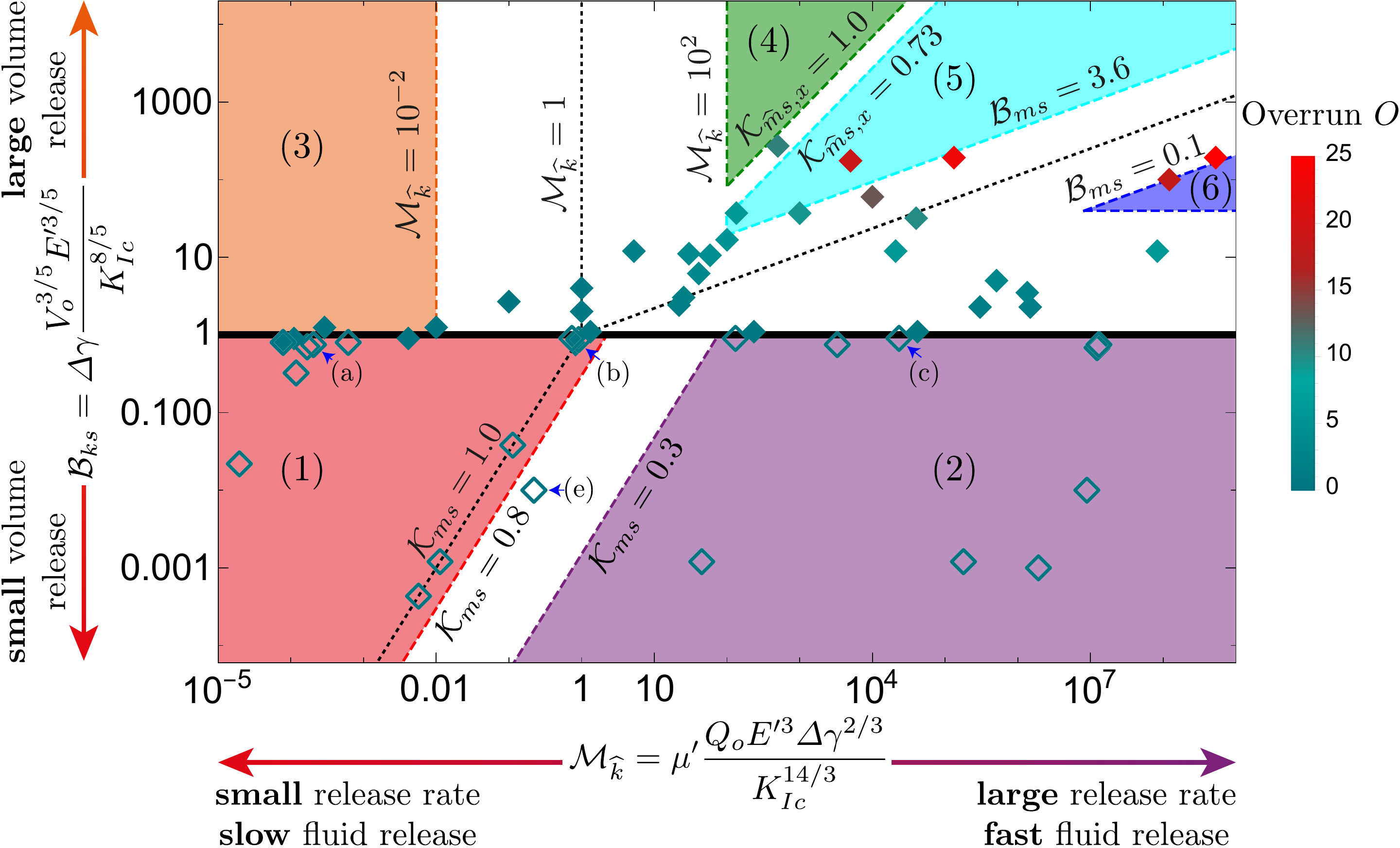}\caption{Structure
of the solution for a finite volume release hydraulic fracture as
a function of the dimensionless buoyancy $\mathcal{B}_{ks}$ (\ref{eq:ShutInBuoyancyK})
and viscosity ${\displaystyle \mathcal{M}_{\widehat{k}}}$ (\ref{eq:MkhatDefinition}).
Each symbol represents a simulation. Arrested fractures have empty
symbols, filled symbols indicate self-sustained buoyancy-driven pulses.
Numbered areas of different colours delimit distinct propagation histories.
The colour of the symbols represents the value of the horizontal overrun
$O$ (\ref{eq:Overrun}). We indicate the simulations presented in
figure \ref{fig:RadialShapes} via blue arrows.\label{fig:ParametricSpace}}
\end{figure}
In the preceding subsections, the necessary and sufficient condition
for the birth of a buoyant fracture $\mathcal{B}_{ks}\geq1$ (see
equation (\ref{eq:ShutInBuoyancyK})) was derived.
The fact that the birth (or not) of a buoyant hydraulic fracture solely
depends on the total released volume and elastic parameters but is
independent of how the volume is accumulated intrinsically derives
from this statement. Furthermore, we discussed that the characteristics
of self-sustained buoyant fractures depend additionally on the dimensionless viscosity $\mathcal{M}_{\widehat{k}}$
(see equation (\ref{eq:MkhatDefinition})), and hence on the specifics of the release (how the volume got released).
These two parameters combined, encompass any possible configuration and thus form the parametric
space of the entire problem (see figure \ref{fig:ParametricSpace}).

First the parametric space can be split into an upper half ($\mathcal{B}_{ks}\geq1$)
where self-sustained buoyant propagation occurs and a lower part ($\mathcal{B}_{ks}<1$)
where the fractures ultimately arrest at depth. We have numerically
investigated this limit, where every symbol in figure \ref{fig:ParametricSpace}
corresponds to a simulation. Empty symbols show simulations where
the fracture ultimately arrests at depth, whereas filled symbols correspond
to cases where self-sustained buoyant growth occurs. In general, figure
\ref{fig:ParametricSpace} shows that the scaling argument that self-sustained
buoyant growth occurs for $\mathcal{B}_{ks}\geq1$ is correct without
any prefactor. Only toughness-dominated fractures at the end of the
release ($\mathcal{K}_{ms}\geq0.8$ where no post-injection radial
propagation occurs) sometimes lead to self-sustained buoyant growth
 for values of $\mathcal{B}_{ks}$ slightly smaller than 1. We use a value of $\mathcal{B}_{ks}=1$
as the limit for the birth of a self-sustained finite volume buoyant
hydraulic fracture. This limit is close to the results obtained in
previous contributions: $\mathcal{B}_{ks}\approx0.90$ for \citet{DaRi20}
and $\mathcal{B}_{ks}\approx0.91$ for \citet{SaZi20}. The equivalent
value of $\mathcal{B}_{ks}$ calculated from the semi-analytically
derived head volume of a propagating toughness-dominated buoyant fracture
by \citet{GaGe14} is significantly higher: $\mathcal{B}_{ks}\approx1.26$.

The parametric space of figure \ref{fig:ParametricSpace} captures
more than the limit between fractures that ultimately arrest and self-sustained
buoyant pulses.We distinguish six well-defined regions, corresponding
to several limiting regimes of radial and buoyant growth: stagnant
fractures with a toughness-dominated end of the release (region 1;
bottom left - red, section \ref{sec:StagnantFractures}), stagnant
fractures with a viscosity-dominated end of the release (region 2;
bottom right - purple, section \ref{sec:StagnantFractures}), toughness-dominated
buoyant fractures at the end of the release (region 3; top left -
orange, section \ref{subsec:ToughDomTrans}), viscosity-dominated
buoyant fractures with a stabilized breadth at the end of the release
(region 4; top centre - dark green, section \ref{subsec:ViscDomBuoyStab}),
viscosity-dominated buoyant fractures without stabilization at the
end of the release (region 5; top centre - light blue, section \ref{subsec:ViscDomBuoyNot}),
and viscosity-dominated radial fractures at the end of the release
(region 6; top right - dark blue, section \ref{subsec:ViscosityDomRadial}).
The distinction between regions 4 and 5 stems from the stagnation
of lateral growth observed for viscosity-dominated buoyant hydraulic
fractures under a constant release rate with a finite toughness \citep{MoLe22c}
and will be detailed later.

\section{Fractures arrested at depth $\mathcal{B}_{ks}<1$\label{sec:StagnantFractures}}

\begin{figure}
\centering{}\includegraphics[width=0.95\textwidth]{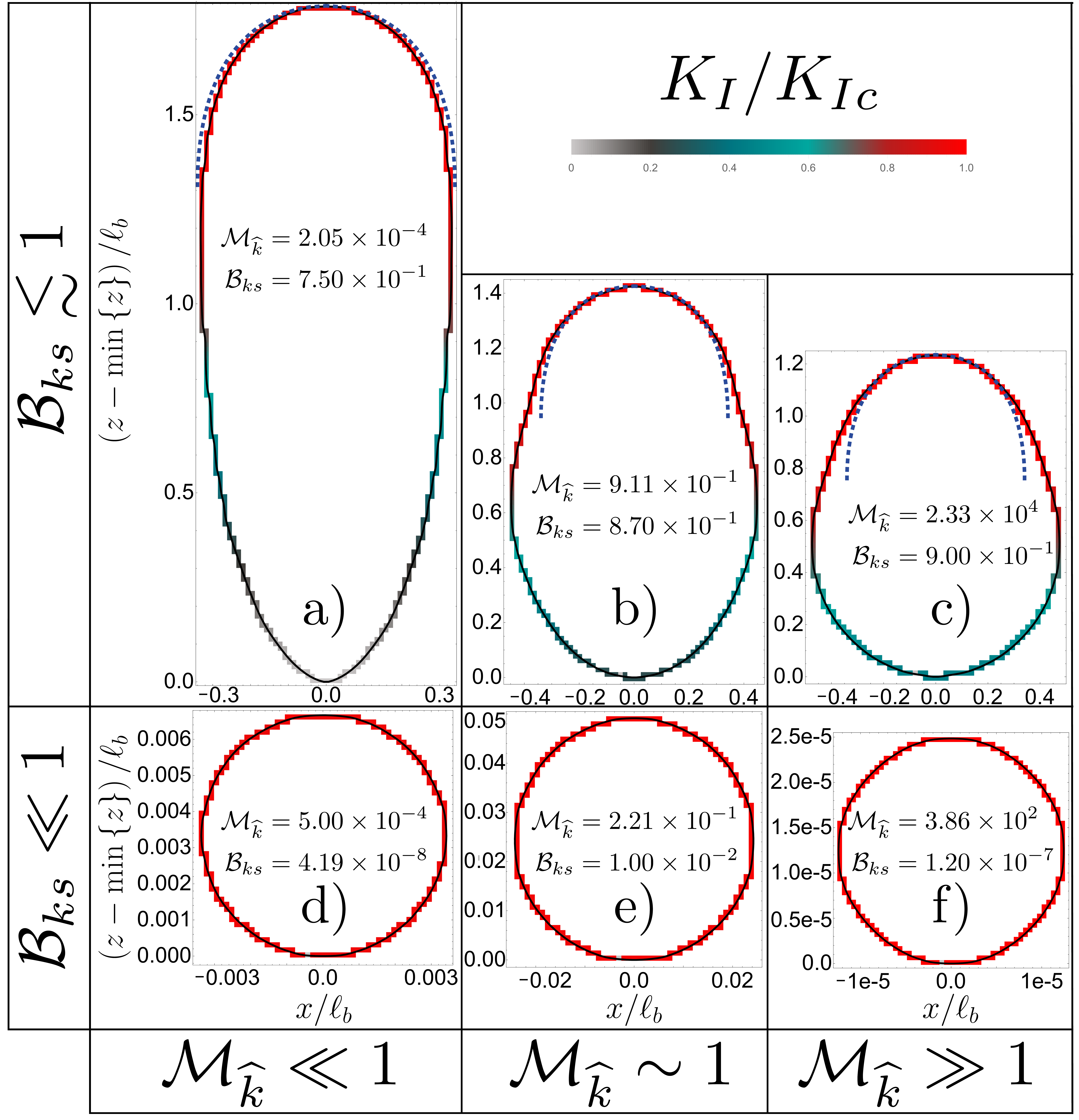}\caption{Final shape and stress intensity factors (SIF) along the front $\mathcal{C}\left(t\right)$
of ultimately arrested fractures at depth ($\mathcal{B}_{ks}<1$)
as a function of $\mathcal{B}_{ks}$ and $\mathcal{M}_{\widehat{k}}$.
Colours indicate the ratio between the local stress intensity factor
$K_{I}$ and the material fracture toughness $K_{Ic}$ from 0 (light
grey) to 1 (red). The blue dashed lines in a) to c) correspond to
the shape of an expanding head of a propagating toughness-dominated
buoyant fracture \citep{GaGe14}.\label{fig:RadialShapes}}
\end{figure} 
Fractures which arrest at depth do not show self-sustained propagation
in the buoyant direction. In the absence of any form of material or
stress heterogeneities and assuming an infinite impermeable elastic
medium, a fracture will arrest only if an insufficient volume is released:
$\mathcal{B}_{ks}<1$.
The lower part of figure \ref{fig:ParametricSpace} distinguishes two propagation histories
for arresting fractures: a region where the fracture is toughness-dominated
at the end of the release (region 1) and one where it is viscosity-dominated
(region 2). As described in section \ref{subsec:FVRadial}, the characteristics
of radially arresting fractures are independent of the propagation
history. In the cases where $\mathcal{B}_{ks}\ll1$, the fracture
has a stress intensity factor (SIF) $K_{I}$ along the entire fracture
front equal to the fracture toughness $K_{Ic}$ (c.f. figures \ref{fig:RadialShapes}d)
- f)). In other words, as long as the final radius of the fracture
$R_{a}$ (\ref{eq:RadialArrestRadius}) is small compared to the buoyancy
length scale $\ell_{b}$ \citep{LiKe91}, the fracture arrests radially,
and the findings obtained in the absence of buoyancy are valid \citep{MoLe21}.

For larger released volumes which are still insufficient for the start
of self-sustained growth ($\mathcal{B}_{ks}\lesssim1$), fracture
elongation occurs before it finally arrests. The fracture footprints
of figures \ref{fig:RadialShapes}a) - c) indicate such elongated
shapes as the dimensionless buoyancy approaches 1. In line with this,
the stress intensity factor is smaller than the material toughness
in the lower part of the fracture. The final elongation of the fracture
is more pronounced for lower values of the dimensionless viscosity
$\mathcal{M}_{\widehat{k}}$. The continuous release case has shown
that toughness- and viscosity-dominated transitions present a distinct
evolution of their shape \citep{MoLe22c}. The different shapes of
the arrested fractures as a function of the dimensionless viscosity
when the released volume, approaches the limiting one is therefore not surprising.

\section{Self-sustained finite volume buoyant fractures: $\mathcal{B}_{ks}>1$\label{sec:BuoyantFractures}}

\subsection{Toughness-dominated, buoyant fractures at the end of the release
(region 3): $\mathcal{M}_{k}\ll1$\label{subsec:ToughDomTrans}}

When the released volume is sufficient to create a buoyant hydraulic
fracture ($\mathcal{B}_{ks}>1$), a set of possible propagation histories
exists as function of the dimensionless viscosity $\mathcal{M}_{k}$.
We first discuss toughness-dominated fractures which, according to
the arguments of sections \ref{subsec:FVRadial} and \ref{sec:FractureStagnation},
must have a transition from radial to buoyant when the release is
still ongoing. This results in a well-established, finger-like buoyant
fracture with a constant volume, toughness-dominated head at the end
of the release. The head characteristics in the case of a continuous
release were obtained from the assumption that $\ell^{\textrm{head}}\left(t\right)\sim b^{\textrm{head}}\left(t\right)$
and elasticity, toughness, and buoyant forces are dominating. If we
additionally restrict these derivations by the finiteness of the total
release volume, the resulting length, opening, and pressure scales
remain unchanged (see equations 4.1 of \citet{MoLe21}) but a time-dependent
dimensionless viscosity emerges
\begin{equation}
\mathcal{M}_{\widehat{k}}^{\left[V\right]}\left(t\right)=\mu^{\prime}\frac{V_{o}E^{\prime3}\varDelta\gamma^{2/3}}{K_{Ic}^{14/3}t}=\mathcal{M}_{\widehat{k}}\frac{t_{s}}{t}.\label{eq:KbarhatHeadViscosity}
\end{equation}
\begin{figure}
\centering{}\includegraphics[width=0.5\textwidth]{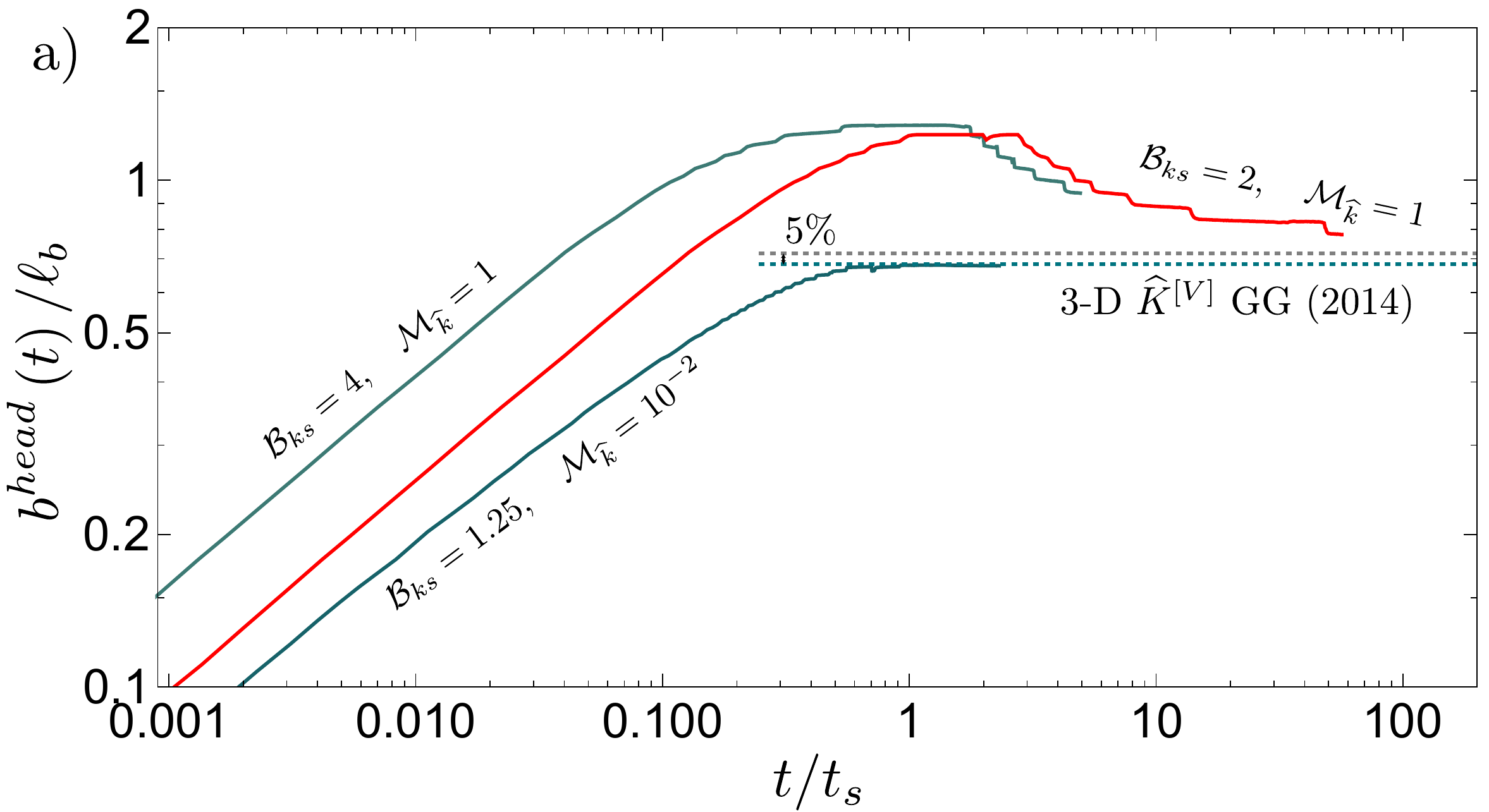}\includegraphics[width=0.5\textwidth]{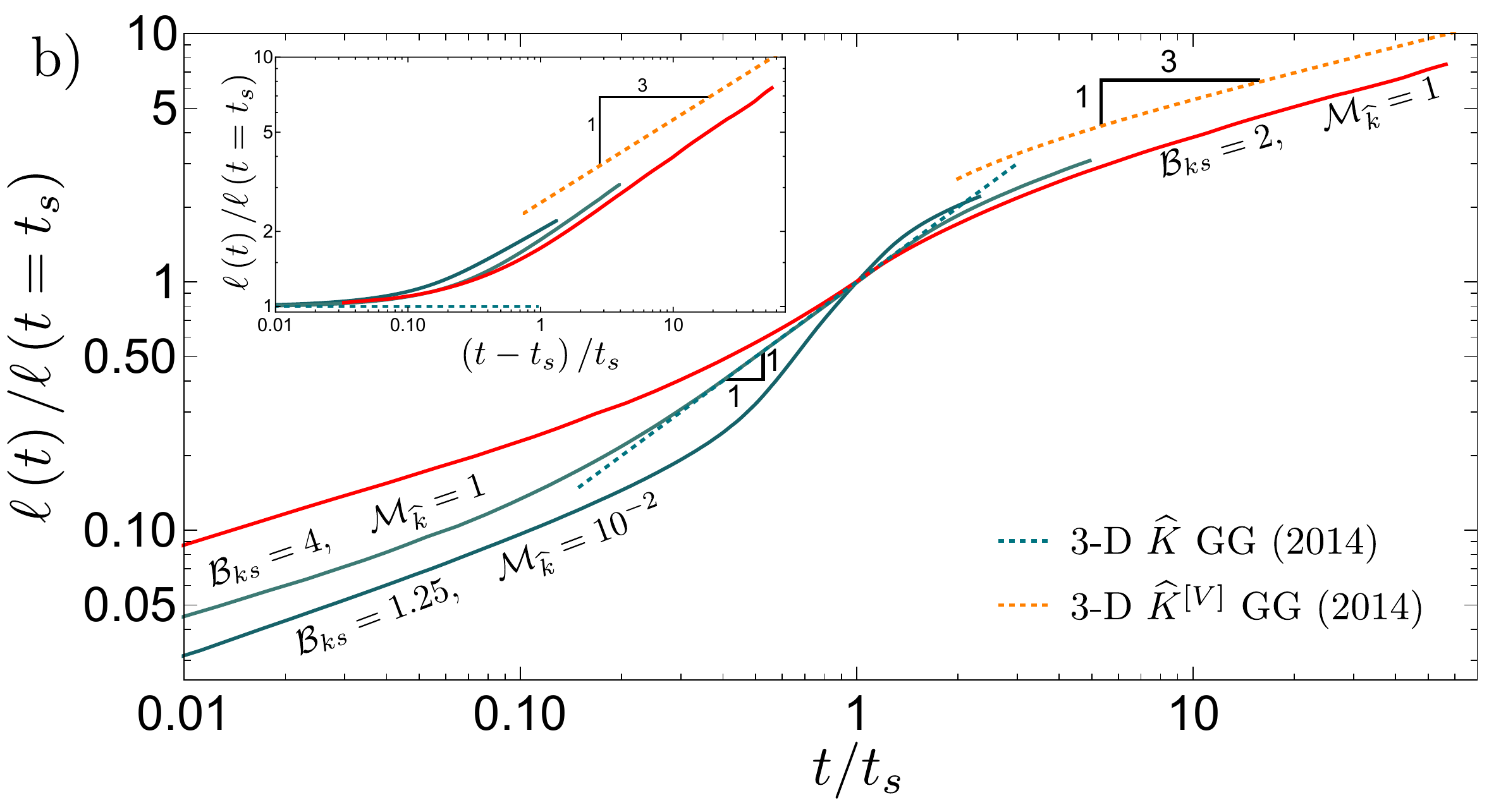}\caption{Toughness-dominated self-sustained buoyant fractures. Evolution of
the dimensionless head breadth $b^{head}\left(t\right)/\ell_{b}$
(a) and fracture length $\ell\left(t\right)/\ell(t=t_{s})$ (b) as
a function of the dimensionless shut-in time $t/t_{s}$. The green-dotted
line corresponds to the limiting 3-D $\widehat{K}$ GG (2014) solution
($b^{head}\left(t\to\infty\right)=\pi^{-1/3}\ell_{b}$ in a), and
the orange dashed line is the 3-D $\widehat{K}^{\left[V\right]}$
GG (2014) solution. The inset of figure (b) shows the same quantity
on the y-axis with a shifted x-axis (e.g. $\left(t-t_{s}\right)/t_{s}$).
\label{fig:ToughnessHeadBreadthConvergence}}
\end{figure}
The decreasing nature of $\mathcal{M}_{\widehat{k}}^{\left[V\right]}$
with time indicates that the fracture head will necessarily become
toughness-dominated at late time. \citet{GaGe14} similarly derived
the finite volume limit and conclude that the head and tail breadth
do not change compared to the continuous release case. Their solution
is thus equivalently representative of any finite volume, buoyant
hydraulic fracture with a finite toughness. We denote their result
hereafter as the 3-D $\widehat{K}^{\left[V\right]}$ GG (2014) solution.
For cases in the intermediate range of $\mathcal{M}_{\widehat{k}}\in\left[10^{-2},10^{2}\right]$,
we check how their head breadth approaches the 3-D $\widehat{K}^{\left[V\right]}$
GG (2014) solution at late time (e.g. $b^{head}\left(t\to\infty\right)=\pi^{-1/3}\ell_{b}$).
We show in figure \ref{fig:ToughnessHeadBreadthConvergence}a) the
evolution of one toughness-dominated simulation with $\mathcal{M}_{\widehat{k}}=10^{-2}$
and two fractures with an intermediate value of $\mathcal{M}_{\widehat{k}}=1$.
The head breadth of the toughness-dominated fracture validates the
limiting solution during the release (dark green line in figure \ref{fig:ToughnessHeadBreadthConvergence}a))
and shows no change after the release has ended. In contrast to this
constant value of the head breadth, the simulations with an intermediate
value of $\mathcal{M}_{\widehat{k}}$ (light green and red lines in
figure \ref{fig:ToughnessHeadBreadthConvergence}a)) have a maximum
value exceeding the limiting breadth at the end of the release. Afterwards,
the head breadth gradually reduces and approaches the limiting 3-D
$\widehat{K}^{\left[V\right]}$ GG (2014) solution. In the continuous
relese case, the limiting breadth is valid for $\mathcal{M}_{\widehat{k}}\leq10^{-2}$,
using equation (\ref{eq:KbarhatHeadViscosity}) we can thus estimate
the time for the fracture to reach the limit as $t\left(\mathcal{M}_{\widehat{k}}^{\left[V\right]}\left(t\right)=10^{-2}\right)=10^{2}\mathcal{M}_{\widehat{k}}t_{s}$
\citep{MoLe22c}. For $\mathcal{M}_{\widehat{k}}\in\left[10^{-2},10^{2}\right]$,
the simulations presented in figure \ref{fig:ToughnessHeadBreadthConvergence}
the 3-D $\widehat{K}^{\left[V\right]}$ GG (2014) solution would be
reached once $t\sim100t_{s}$. From the rate with which the breadth
approaches the 3-D $\widehat{K}^{\left[V\right]}$ GG (2014) observed
in figure \ref{fig:ToughnessHeadBreadthConvergence}a), this estimate
seems reasonable. In fact, the fracture with $\mathcal{M}_{\widehat{k}}=1$
and $\mathcal{B}_{ks}=2$ is already within $15\%$ of the limiting
solution at $t\sim50t_{s}$.

We derive the scaling of the viscosity-dominated tail of such a late-time
solution using the assumption of a constant frature breadth on the
order of the breadth of the head $b\sim\ell_{b}=K_{Ic}^{2/3}/\varDelta\gamma^{2/3}$
as
\begin{align}
\ell_{\widehat{k}}^{\left[V\right]}\left(t\right)=\frac{V_{o}^{2/3}\varDelta\gamma^{7/9}\,t^{1/3}}{K_{Ic}^{4/9}\mu^{\prime1/3}},\quad b_{\widehat{k}}^{\left[V\right]}=\frac{K_{Ic}^{2/3}}{\varDelta\gamma^{2/3}}\equiv\ell_{b}\label{eq:KhatVScales}\\
w_{\widehat{k}}^{\left[V\right]}\left(t\right)=\frac{V_{o}^{1/3}\mu^{\prime1/3}}{K_{Ic}^{2/9}\varDelta\gamma^{1/9}t^{1/3}},\quad p_{\widehat{k}}^{\left[V\right]}\left(t\right)=E^{\prime}\frac{\varDelta\gamma^{5/9}V_{o}^{1/3}\mu^{\prime1/3}}{t^{1/3}K_{Ic}^{8/9}}.
\end{align}
where we use $\widehat{\cdot}$ to refer to a buoyant scaling. These
scales are obtained from the continuous release scales by replacing
$Q_{o}$ with $V_{o}/t$, and reveal a sub-linear growth of the fracture
height according to a power law of the form $\ell\left(t\right)\sim t^{1/3}$.
Note that these scales have been obtained by \citet{GaGe14} when
deriving their 3-D $\widehat{K}^{\left[V\right]}$ GG (2014) solution.
We present in figure \ref{fig:ToughnessHeadBreadthConvergence}b)
the evolution of dimensionless fracture length $\ell\left(t\right)/\ell\left(t=t_{s}\right)$
as a function of the dimensionless time $t/t_{s}$. The dark green
line with a $1:1$ slope indicates the scaling derived temporal power-law
for a toughness-dominated buoyant hydraulic fracture under a continuous
fluid release. The two simulations with low $\mathcal{B}_{ks}$ (dark
red and green) cannot reach this intermediate regime, as they are
not propagating long enough in this $\widehat{K}$-regime (see the
discussion in section 4.4 of \citet{MoLe22c}). The simulation with
$\mathcal{B}_{ks}=4$ reaches this limit for about one order of magnitude
in time before decelerating towards the late-time power law predicted
by the scaling of equation (\ref{eq:KhatVScales}). A similar deceleration
is observed for the other two simulations without any of the simulations
reaching the limitting $\ell\left(t\right)\sim t^{1/3}$ power-law.
The orange dashed line indicates the 3-D $\widehat{K}^{\left[V\right]}$
GG (2014) for fracture length, which we would expect to be valid at
late times. The inset of figure \ref{fig:ToughnessHeadBreadthConvergence}b)
sets the time when the release ends as zero according to the hypothesis
of \citet{GaGe14}. This correction of the data highlights the tendency
of the fracture length of all simulations to approach the limiting
solution. A late-time validation of the solution can be expected as
the relative difference between the predicted length and the simulation
with $\mathcal{B}_{ks}=2$ and $\mathcal{M}_{\widehat{k}}=1$ at the
end of the simulation is only on the order of $23\%$. These findings
indicate that buoyant fractures with a finite toughness will have
a late-time behaviour akin to the 3-D $\widehat{K}^{\left[V\right]}$
GG (2014) solution. Even though this late-time behaviour will be consistent,
it also shows that the exact shape of the fracture will depend on
both parameters, $\mathcal{M}_{\widehat{k}}$ and $\mathcal{B}_{ks}$.
Only the breadth close to the head, the head itself, and the growth
rate will be equivalent to the 3-D $\widehat{K}^{\left[V\right]}$
GG (2014) solution. To get an idea of the overall fracture shape,
we define a shape parameter called the overrun as
\begin{equation}
O=\frac{\underset{z,t}{\text{max}}\left\{ b\left(z,t\right)\right\} -\pi^{-1/3}\ell_{b}}{\pi^{-1/3}\ell_{b}}.\label{eq:Overrun}
\end{equation}
This parameter defines how much the maximum lateral extent exceeds
the late-time head breadth $\pi^{-1/3}\ell_{b}$. $O$ has a lower
bound of $0$, reached for fully toughness-dominated fractures with
$\mathcal{M}_{\widehat{k}}\leq10^{-2}$. This limit is validated by
the simulation reported in this section with $\mathcal{M}_{\widehat{k}}=10^{-2}$
and $\mathcal{B}_{ks}=1.25$ which effectively has an overrun of $0.$
For the fractures in between the toughness and viscosity dominated
limit of the continuous release with a uniform breadth (e.g. $\mathcal{M}_{\widehat{k}}\in\left[10^{-2},10^{2}\right]$),
the overrun cannot be predicted by scaling laws. From the observation
of figure 8 of \citet{MoLe22c}, we can however derive that it will
increase with increasing values of $\mathcal{M}_{\widehat{k}}$.The
overrun of the two simulations reported here is respectively $0.88$
($\mathcal{M}_{\widehat{k}}=1$ and $\mathcal{B}_{ks}=4$) and $0.80$
($\mathcal{M}_{\widehat{k}}=1$ and $\mathcal{B}_{ks}=2$). We display
the value of the overrun for simulations which lead to a buoyant hydraulic
fracture in figure \ref{fig:ParametricSpace}. Within the region of
the toughness-dominated fractures with a buoyant end of the release
(region 1), the values are effectively $0$. The overrun increases
with the value of $\mathcal{M}_{\widehat{k}}$ towards the viscosity-dominated
domain (regions 4 to 6) and will be estimated using scaling arguments
later.

\subsubsection{Numerical Limitations\label{subsec:NumericalLimitations}}

The fact that no simulations propagating for longer times - which
would ultimately exhibit the 3-D $\widehat{K}^{\left[V\right]}$ GG
(2014) solution - are reported deserves discussion. These simulations
have multiple numerical challenges: their overall computational cost
and the numerical treatment of closing cells at the bottom of the
fracture, among others. We illustrate the computational cost by the
example of a toughness-dominated buoyant hydraulic fracture. Such
fractures accelerate around the transition from radial to buoyant
before slowing down to the ultimately constant velocity. \citet{MoLe22c}
report that for their simulations, the acceleration terminates at
a dimensionless time of approximately $t/t_{k\widehat{k}}\approx3$,
where $t_{k\widehat{k}}$ is the transition time from radial to buoyant
(see equation 3.6 of \citet{MoLe22c}). Observation of figure \ref{fig:ToughnessHeadBreadthConvergence}b)
shows that after the end of the release, additional time is required
to transition to the late-time buoyant pulse solution. This figure
gives an estimate of the time to reach the 3-D $\widehat{K}^{\left[V\right]}$
GG (2014) solution of $t\sim100t_{s}$. An estimate of the fracture
extent for a simulation with $\mathcal{M}_{\widehat{k}}=10^{-2}$
at this time, based on a growth according to the power law of equation
(\ref{eq:KhatVScales}), gives $\ell\sim1600\ell_{b}$. The computational
cost can now be estimated by taking a discretization of approximately
$44$ elements per $\ell_{b}$ (see section 4.2 of \citet{MoLe22c})
and an approximation of the constant breadth of $b\left(t\right)\approx\pi^{-1/3}\ell_{b}$,
yielding about $2\times10^{6}$ elements in the fracture. Our current
implementation of PyFrac \citep{ZiLe20} is able to handle buoyant
simulations covering up to $20$ orders of magnitude in time and up
to $15$ orders of magnitude in space within about $4$ weeks of computation
time on a multithreaded Linux desktop system with twelve Intel\textregistered
Core i7-8700 CPUs, using at most 30 GB of RAM, and arising to a discretization
of up to $2\times10^{5}$ elements within the fracture footprint.

An additional issue presents closing cells at the bottom of the fracture.
As the opening continuously reduces (see $w_{\widehat{k}}^{\left[V\right]}$
in equation (\ref{eq:KhatVScales})) and we do not allow for complete
fracture healing, a minimum width activates \citep{ZiLe20}. From
this arise two effects: First, elastic contact stress changes the
stress distribution and the overall behaviour. Second, some volume
gets trapped, reducing the one available for fracture propagation.
The latter will artificially slow down propagation \citep{Pezz22}
and ultimately arrest the fracture. Both phenomena increase the non-linearity
of the system, such that convergence is challenging, which leads to
a breakdown of the simulation at late time $t\gg t_{s}$.

\subsection{Viscosity-dominated at the end of the release (regions 4 to 6): $\mathcal{M}_{k}\gg1$}

\begin{figure}
\centering{}\includegraphics[width=0.49\textwidth]{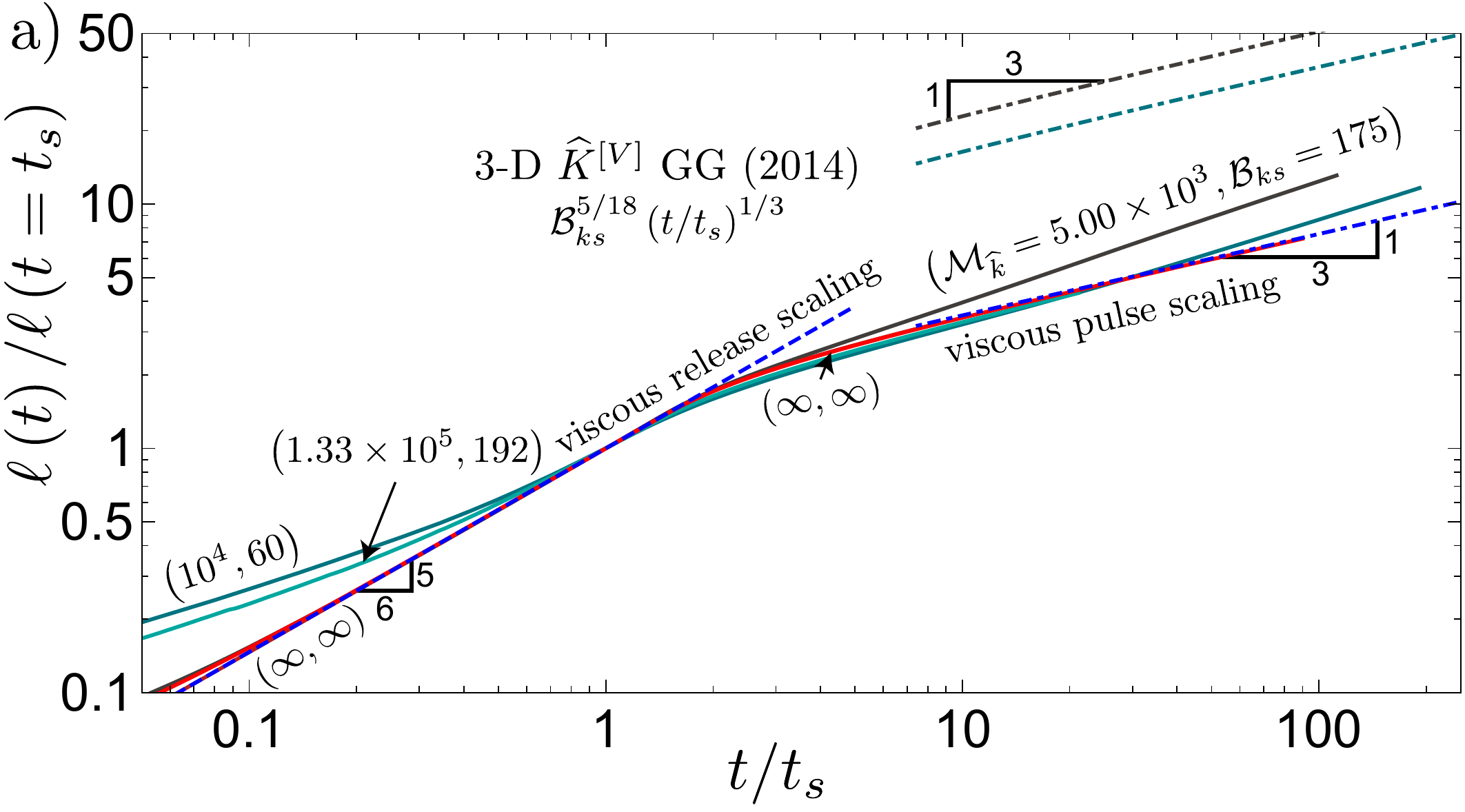}\enskip{}\includegraphics[width=0.49\textwidth]{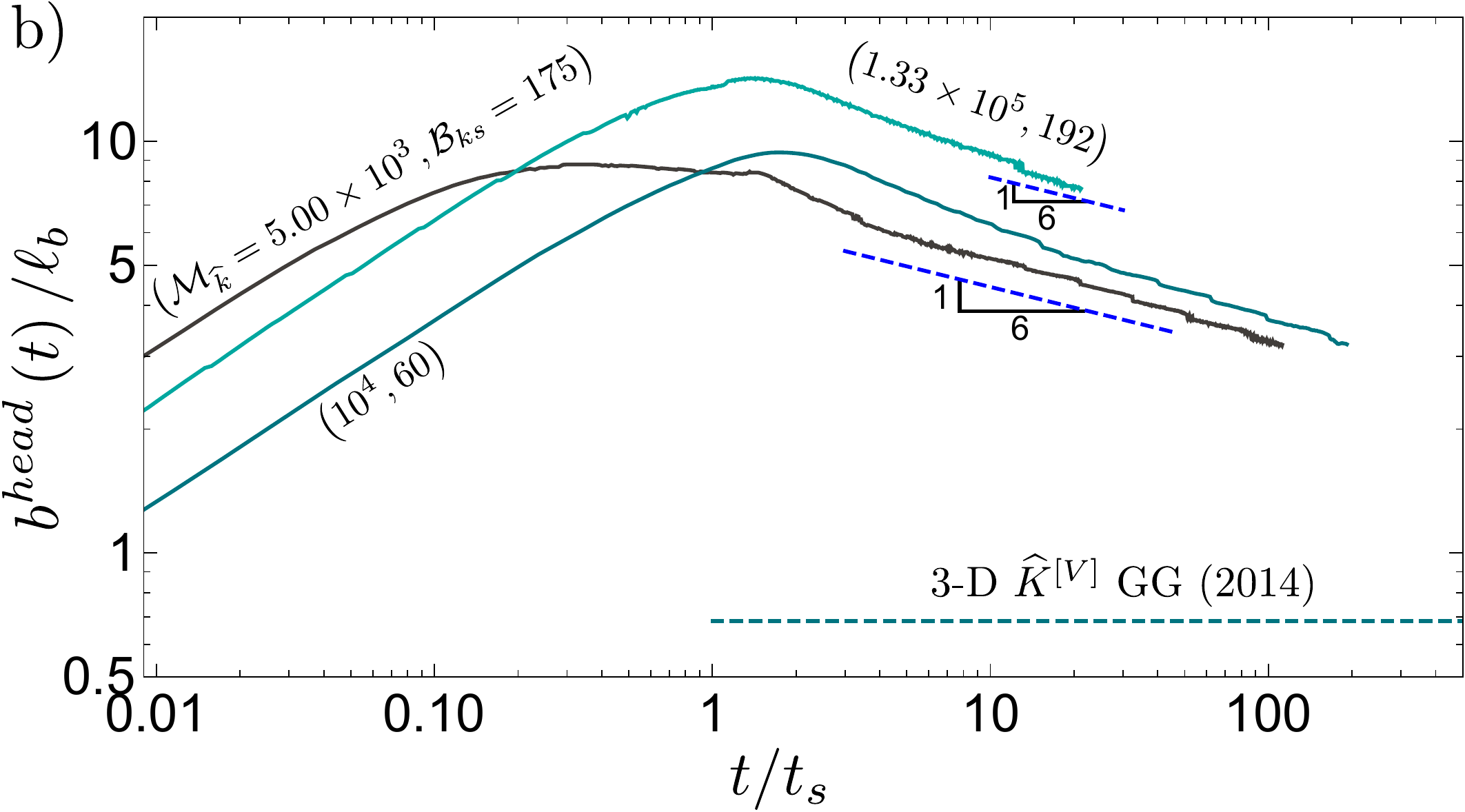}\caption{Evolution of fracture length $\ell\left(t\right)/\ell(t=t_{s})$ (a)
and head breadth $b^{head}\left(t\right)/\ell_{b}$ for viscosity-dominated
buoyant non-stabilized fractures at the end of the release as a function
of the dimensionless shut-in time $t/t_{s}$: $\mathcal{M}_{\widehat{k}}\gg1,\,\mathcal{B}_{ks}\protect\geq1,\,\mathcal{K}_{\widehat{m}x,s}<1$.
Coloured dashed-dotted lines in (a) show the corresponding late-time,
3-D $\widehat{K}^{\left[V\right]}$ GG (2014) solution, the blue dashed
line the continuous release buoyant scaling $\left(\ell\left(t\right)/\ell(t=t_{s})\sim t^{5/6}\right)$,
the blue dashed-dotted line a numerical zero-toughness fit $\ell\left(t\right)/\ell\left(t=t_{s}\right)\approx1.62\left(t/t_{s}\right)^{0.33}$
(matching the $\widehat{M}^{\left[V\right]}$-scaling). The green
dashed line in (b) indicates the late-time limit of the 3-D $\widehat{K}$
GG (2014) solution for the corresponding simulation. The blue-dashed
line indicates the scaling dependence in the $\widehat{M}^{\left[V\right]}$-scaling,
$b^{head}\left(t\right)\sim t^{-1/6}$. Note that the two zero-toughness
simulations differ by their value of $\mathcal{B}_{ms}$ ($100$ for
the dark red and $25$ for the light red simulation).\label{fig:ViscousDuring}}
\end{figure}
The difference between a buoyant or radial end of the release has
been shown to depend on the dimensionless viscosity at the end of
the release $\mathcal{B}_{ms}$ (see equation (\ref{eq:ShutInBuoyancyM}),
section \ref{subsec:ViscTransLimitDeriv}). For an accurate evaluation
of the emerging shape, an additional separation between two possible
cases of buoyant fractures at the end of the release is required.
\citet{MoLe22c} have shown that whenever a finite fracture toughness
is present (e.g. $K_{Ic}\neq0$), lateral growth stabilizes within
a finite time at $\underset{z,t}{\text{max}}\left\{ b\left(z,t\right)\right\} \propto\mathcal{M}_{\widehat{k}}^{2/5}\ell_{b}$.
The time of stabilization is related to a dimensionless lateral toughness
$\mathcal{K}_{\widehat{m},x}(t)$ (see their equation 6.1), which
we can evaluate at the end of the release 
\begin{equation}
\mathcal{K}_{\widehat{m}s,x}=\mathcal{K}_{\widehat{m},x}(t=t_{s})=K_{Ic}\frac{\varDelta\gamma^{1/8}\,t_{s}^{1/3}}{E^{\prime19/24}V_{o}^{1/8}\mu^{\prime1/3}}=\mathcal{M}_{\widehat{k}}^{1/3}\mathcal{B}_{ks}^{25/72}.\label{eq:MhatTailTough}
\end{equation}
A value of $\mathcal{K}_{\widehat{m}s,x}\geq1$ indicates that lateral
growth has ceased, whereas a value below one means that the fracture
is still growing laterally as $b\sim t^{1/4}$ (see equation 5.4 of
\citet{MoLe22c}).

\subsubsection{Viscosity-dominated, buoyant fracture at the end of the release without
laterally stabilized breadth (region 5): $\mathcal{B}_{ms}\protect\geq1$
and$\mathcal{K}_{\widehat{m}s,x}\ll1$\label{subsec:ViscDomBuoyNot}}

We consider first the case of zero-fracturing toughness by the development
of a tail scaling. The principle hypotheses are buoyant forces, elasticity
and viscous energy dissipation at first order and an aspect ratio
scaling like the respective lateral and horizontal fluid velocities
($\ell\left(t\right)/b\left(t\right)\sim v_{z}\left(t\right)/v_{x}\left(t\right)$)
\begin{align}
\ell_{\widehat{m}}^{\left[V\right]}=\frac{V_{o}^{1/2}\varDelta\gamma^{1/2}\,t^{1/3}}{E^{\prime1/6}\mu^{\prime1/3}},\quad b_{\widehat{m}}^{\left[V\right]}=\frac{E^{\prime1/4}V_{o}^{1/4}}{\varDelta\gamma^{1/4}}\label{eq:VMhatScaling}\\
w_{\widehat{m}}^{\left[V\right]}=\frac{V_{o}^{1/4}\mu^{\prime1/3}}{E^{\prime1/12}\varDelta\gamma^{1/4}t^{1/3}},\quad p_{\widehat{m}}^{\left[V\right]}=\frac{E^{\prime2/3}\mu^{\prime1/3}}{t^{1/3}}.\nonumber 
\end{align}
Note that \citet{DaRi23} presented the same scaling for fracture
length. The finite volume inherently prevents the infinite lateral
growth observed for a continuous release and $b_{\widehat{m}}^{\left[V\right]}$
is time-independent. Figure \ref{fig:ViscousDuring}c) shows limited
lateral growth for all simulations. It is interesting to note, that
the scaling predicts a fracture length evolution with a $\ell\sim t^{1/3}$
power-law, equivalent to the height evolution in the toughness-dominated
case. Figure \ref{fig:ViscousDuring}a) shows this evolution for various
viscosity-dominated simulations. When $\mathcal{M}_{\widehat{k}}$
is sufficiently large and $\mathcal{K}_{\widehat{m}s}\ll1$ (in other
words when the fracture is sufficiently far from lateral stabilization)
the $1:3$ slope predicted by the scaling of equation (\ref{eq:VMhatScaling})
emerges. However, the height growth quickly departs from the $t^{1/3}$
power-law. The reason is the time-dependent inflow rate of the head
(derived from the scaling of equation (\ref{eq:VMhatScaling}))
\begin{align}
\ell_{\hat{m}}^{head,\left[V\right]}=b_{\hat{m}}^{head,\left[V\right]}=\frac{E^{\prime11/24}V_{o}^{1/8}\mu^{\prime1/6}}{\varDelta\gamma^{5/8}t^{1/6}},\hspace{1em}w_{\hat{m}}^{head,\left[V\right]}=\frac{V_{o}^{1/4}\mu^{\prime1/3}}{E^{\prime1/12}\varDelta\gamma^{1/4}t^{1/3}}\label{eq:VMhatHeadScaling}\\
p_{\hat{m}}^{head,\left[V\right]}=\frac{E^{\prime11/24}V_{o}^{1/8}\mu^{\prime1/6}\varDelta\gamma^{3/8}}{t^{1/6}},\hspace{1em}V_{\hat{m}}^{head,\left[V\right]}=\frac{E^{\prime5/6}V_{o}^{1/2}\mu^{\prime2/3}}{\varDelta\gamma^{3/2}t^{2/3}},\nonumber 
\end{align}
revealing a shrinking viscous head.

Considering now a finite fracture toughness, a dimensionless toughness
can be obtained in the head
\begin{equation}
\mathcal{K}_{\widehat{m}}^{[V]}\left(t\right)=K_{Ic}\frac{t^{1/4}}{E^{\prime11/16}V_{o}^{3/16}\varDelta\gamma^{1/16}\mu^{\prime1/4}}=\mathcal{B}_{ks}^{5/48}\mathcal{M}_{\widehat{k}}^{[V]}\left(t\right)^{-1/4}=\mathcal{B}_{ks}^{5/48}\mathcal{M}_{\widehat{k}}^{-1/4}\left(\frac{t}{t_{s}}\right)^{1/4}.\label{eq:VMhatDimToughDef}
\end{equation}
Equation (\ref{eq:VMhatDimToughDef}) indicates that the head will
become toughness dominated at late times as $\mathcal{K}_{\widehat{m}}^{[V]}\left(t\right)$
increases with time. From this observation, we anticipate that the
region close to the propagating head will ultimately follow the 3-D
$\widehat{K}^{\left[V\right]}$ GG (2014) head solution (see section
\ref{subsec:NumericalLimitations}) and derive the characteristic
time scale of the transition

\begin{equation}
t_{\widehat{m}\widehat{k}}^{[V]}=\frac{E^{\prime11/4}V_{o}^{3/4}\varDelta\gamma^{1/4}\mu^{\prime}}{K_{Ic}^{4}}.\label{eq:VMhatToKhatTimeScale}
\end{equation}
Evaluating the viscosity-dominated head scaling (see equations (\ref{eq:VMhatHeadScaling}))
at this characteristic time gives the scales of the toughness-dominated
head (see (\ref{eq:KhatVScales})). This observation implies that
even though the shape further away from the head varies, the length
scale $\ell\left(t\right)_{\widehat{k}}^{\left[V\right]}$ becomes
applicable. Relating the two length scales of buoyant fractures from
a finite volume release
\begin{equation}
\ell{}_{\widehat{k}}^{\left[V\right]}\left(t\right)=\mathcal{B}_{ks}^{5/18}\ell{}_{\widehat{m}}^{\left[V\right]}\left(t\right)\label{eq:RelationBetweenBuoyantLengths}
\end{equation}
shows that $\ell{}_{\widehat{k}}^{\left[V\right]}\left(t\right)\geq\ell{}_{\widehat{m}}^{\left[V\right]}\left(t\right)$
for a buoyant fracture (as $\mathcal{B}_{ks}\geq1$). The observation
of figure \ref{fig:ViscousDuring}a) shows the fracture deviation
from the lower, viscosity-dominated solution towards the upper, toughness-dominated
solution (shown by dashed-dotted lines for two simulations). The observed
faster growth in height originates in the narrowingof the tail, creating
a lateral inflow from the stagnant parts of the fracture into a central
tube of the fixed breadth predicted by the 3-D $\widehat{K}^{\left[V\right]}$
GG (2014) solution. We do not present a simulation that finishes the
transition to the toughness-dominated regime due to its excessive
computational cost (see the discussion in section \ref{subsec:NumericalLimitations}).

In equation (\ref{eq:Overrun}), we have introduced the overrun as
a characteristic of the fracture shape. In the case of viscous fractures
with a buoyancy-dominated, laterally non-stabilized end of the release,
such overrun can be estimated from the viscous scaling as
\begin{equation}
O_{\widehat{m}}=\frac{b_{\widehat{m}}^{\left[V\right]}-\pi^{-1/3}\ell_{b}}{\pi^{-1/3}\ell_{b}}=\pi^{1/3}\frac{E^{\prime1/4}V_{o}^{1/4}\varDelta\gamma^{5/12}}{K_{Ic}^{2/3}}-1=\pi^{1/3}\mathcal{B}_{ks}^{5/12}-1.\label{eq:OverrunViscBuoyNotStab}
\end{equation}
The increase of the overrun with the value of the dimensionless buoyancy
$\mathcal{B}_{ks}$ is observable in figure \ref{fig:ParametricSpace}.

\subsubsection{Viscosity-dominated, buoyant fracture at the end of the release with
laterally stabilized breadth (region 4): $\mathcal{B}_{ms}\protect\geq1$
and $\mathcal{K}_{\widehat{m}s,x}\protect\geq1$\label{subsec:ViscDomBuoyStab}}

Lateral stabilization of buoyant, viscosity-dominated fractures occurs
when the volume of the fracture head becomes constant, leading to
two fixed points, the laterally stabilized breadth of $\underset{z,t}{\text{max}}\left\{ b\left(z,t\right)\right\} \sim\mathcal{M}_{\widehat{k}}^{2/5}\ell_{b}$
and the constant volume, constant breadth head. The section of extending
fracture breadth in between the two conserves its shape, creating
a fracture where elongation concentrates within the zone of laterally
stabilized breadth. From this observation, one can draw an analogy
to a toughness-dominated buoyant fracture (see section \ref{subsec:ToughDomTrans}).
The scales of this equivalent toughness-dominated fracture are related
through a factor of $\mathcal{M}_{\widehat{k}}^{2/5}$, such that
the behaviour after the end of the release will be the same as presented
in section \ref{subsec:ToughDomTrans}, differing only by the starting
point ($\mathcal{M}_{\widehat{k}}^{2/5}$ instead of $\mathcal{M}_{\widehat{k}}$).

Because the processes after the end of the release do not differ from
toughness-dominated fractures, we omit a detailed discussion of this
case hereafter and only list the difference in the shape parameter
\begin{equation}
O_{\widehat{m}}^{stab}=\frac{\mathcal{M}_{\widehat{k}}^{2/5}\ell_{b}-\pi^{-1/3}\ell_{b}}{\pi^{-1/3}\ell_{b}}=\pi^{1/3}\mathcal{M}_{\widehat{k}}^{2/5}-1.\label{eq:OverViscBuoyStab}
\end{equation}
The overrun in the non-stabilized case of viscosity-dominated fractures
depends solely on the dimensionless buoyancy $\mathcal{B}_{ks}$ and,
as such, on the total released volume and elastic parameters. In contrast,
the governing parameter of the stabilized case is the dimensionless
viscosity $\mathcal{M}_{\widehat{k}}$, and the history of the release
(how the total volume gets accumulated) governs the overrun of the
fracture.

\subsubsection{Viscosity-dominated fracture with negligible buoyancy at the end
of the release (region 6): $\mathcal{B}_{ms}\ll1$ \label{subsec:ViscosityDomRadial}}

\begin{figure}
\centering{}\includegraphics[width=\textwidth]{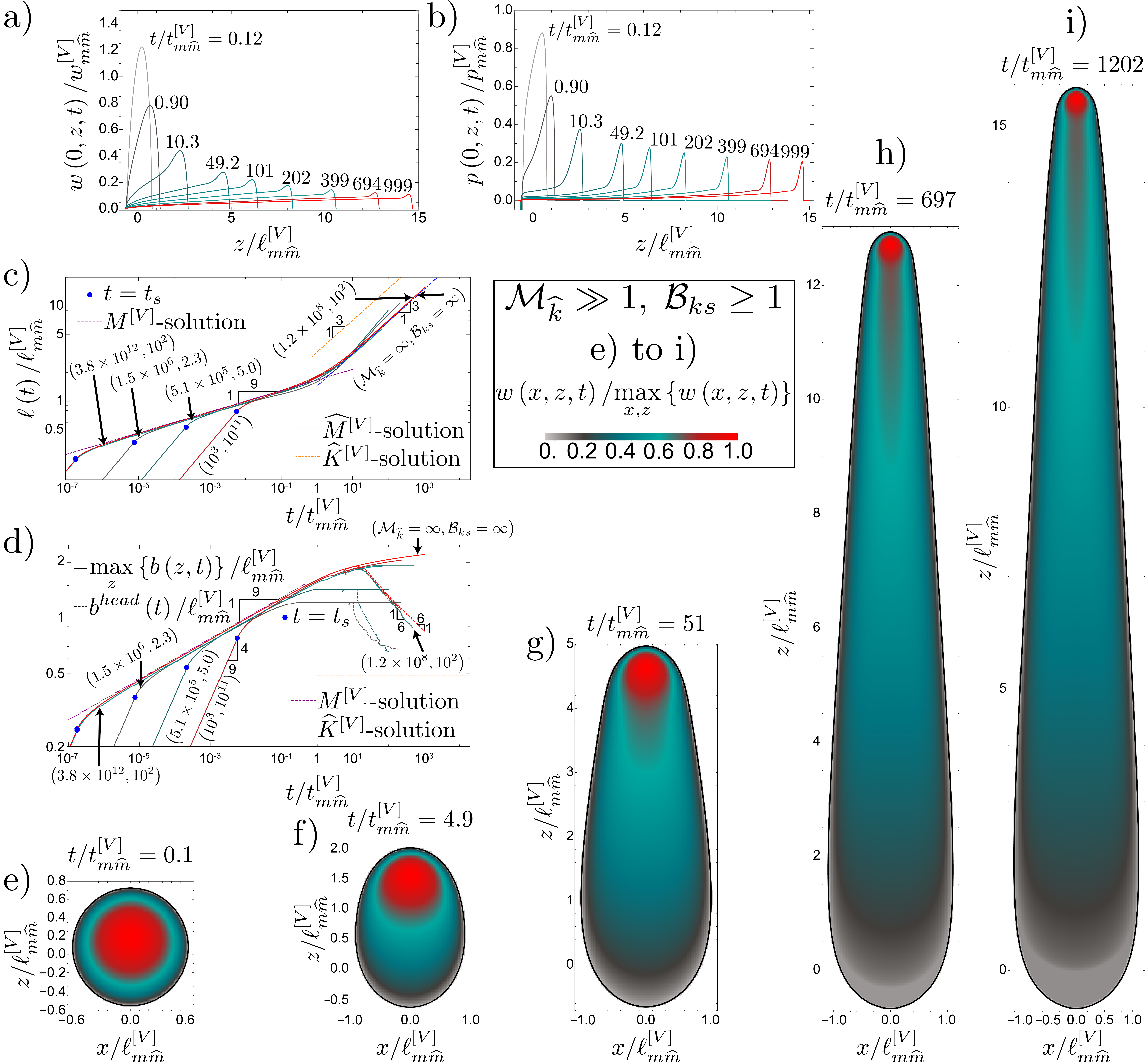}\caption{Viscosity-dominated, fractures with negligible buoyancy at the end
of the release: $\mathcal{B}_{ks}\protect\geq1$ and $\mathcal{B}_{ms}\ll1$.
a) Opening along the centerline $w\left(0,z,t\right)/w_{m\widehat{m}}^{\left[V\right]}$
for $\mathcal{M}_{\widehat{k}}=\infty,\,\mathcal{B}_{ks}=\infty,$
and $\mathcal{B}_{ms}=10^{-3}$ (zero-toughness case). b) Net pressure
along the centerline $p\left(0,z,t\right)/p_{m\widehat{m}}^{\left[V\right]}$
for the same case as in a). c) Fracture length $\ell\left(t\right)/\ell_{m\widehat{m}}^{\left[V\right]}$
for large viscosity $\mathcal{M}_{\widehat{k}}\in\left[5.1\times10^{5},\infty\right]$
simulations. The blue dashed line is a fit of the zero-toughness simulation
$\ell\left(t\right)\propto t^{0.33}$. d) Fracture breadth $b\left(t\right)/\ell_{m\widehat{m}}^{\left[V\right]}$
(continuous lines) and head breadth $b^{head}\left(t\right)/\ell_{m\widehat{m}}^{\left[V\right]}$
(dashed lines) for the same simulations. Purple dashed lines indicate
the $M^{\left[V\right]}$-solution \citep{MoLe21}, orange dashed
lines the 3-D $\widehat{K}^{\left[V\right]}$ GG (2014) solution for
the highest value of $\mathcal{B}_{ks}$. e - i) Evolution of the
fracture footprint from radial e) towards the late time shape h) and
i)) for the zero-toughness simulation. For the definition of the transition scales $\cdot_{m\widehat{m}}^{\left[V\right]}$
see table \ref{Apptab:TransitionScales}.\label{fig:BuoyantAfter}}
\end{figure}

This type of fracture becomes buoyant in the pulse propagation phase
as long as its dimensionless buoyancy $\mathcal{B}_{ks}$ (see (\ref{eq:ShutInBuoyancyK}))
is larger than one. This transition from radial to buoyant propagation
is characterized by the dimensionless buoyancy of the viscous pulse
$M^{\left[V\right]}$-scaling $\mathcal{B}_{m}^{\left[V\right]}\left(t\right)$
(see equation (\ref{eq:BmPulse})) and has a characteristic transition
time
\begin{equation}
t_{m\widehat{m}}^{\left[V\right]}=\frac{E^{\prime5/4}\mu^{\prime}}{V_{o}^{3/4}\varDelta\gamma^{9/4}}=\mathcal{B}_{ks}^{-5/2}t_{\widehat{m}\widehat{k}}^{\left[V\right]}.\label{eq:ViscPulseRadialToBuoyantTransitionTime}
\end{equation}
The corresponding transition length scale is equivalent to the constant
breadth of a buoyant viscosity-dominated fracture $\ell_{m}^{\left[V\right]}\left(t=t_{m\widehat{m}}^{\left[V\right]}\right)=\ell_{m\widehat{m}}^{\left[V\right]}=b_{\widehat{m}}^{\left[V\right]}$,
indicating that the maximum breadth is reached at the transition.
Figure \ref{fig:BuoyantAfter}d) shows that for an increasing dimensionless
buoyancy $\mathcal{B}_{ks}$ (see (\ref{eq:ShutInBuoyancyK})), the
growth of the maximal breadth continues (continuous lines) after transition
but remains within the order of magnitude predicted by the scaling
(\ref{eq:VMhatScaling}). Lateral growth ultimately tapers off at
about $3\ell_{m\widehat{m}}^{\left[V\right]}$ at $t\approx10^{3}t_{m\widehat{m}}^{\left[V\right]}$.
The expected overrun becomes equivalent to the case of a non-stabilized,
buoyant viscosity-dominated end of the release (see equation (\ref{eq:OverrunViscBuoyNotStab})).

The scaling for these fractures is given by equations (\ref{eq:VMhatScaling})
and (\ref{eq:VMhatHeadScaling}). Despite the distinct propagation
histories, the late-time fracture footprint does not vary significantly
(see figure \ref{fig:DifferentBuoyantShapes}). Similar to the case
of a constant release, the fracture first becomes somewhat elliptical
with a peak in pressure and opening appearing in the fracture head.
Propagation then deviates to the buoyant direction with a continuously
shrinking head, and no saddle point develops between the maximum lateral
extent and the head. In the case of finite fracture toughness, an
inflexion point forms in this area, such that the evolution of the
breadth towards the head becomes convex at the transition time $t_{\widehat{m}\widehat{k}}^{[V]}$
(see equation (\ref{eq:VMhatToKhatTimeScale})). Note that the bottom
end of the fractures in figure \ref{fig:BuoyantAfter}h), i) seem
to be of uniform opening. This is a direct consequence of the numerical
scheme where the minimum width was activated.

When observing the evolution of the fracture length and head breadth,
one observes that the simulations approach the 3-D $\widehat{K}^{\left[V\right]}$
GG (2014) solution for cases with a finite toughness. The breadth
and length evolution of the 3-D $\widehat{K}^{\left[V\right]}$ GG
(2014) in the viscous buoyant scaling (see equations (\ref{eq:VMhatScaling})
and (\ref{eq:VMhatHeadScaling})) depends on the value of $\mathcal{B}_{ks}$
such that we only indicate one of the possible late-time solutions.
We pick the one which is most likely to be reached, corresponding
to the smallest value of $\mathcal{B}_{ks}$ for the length and the
largest for the breadth with dashed orange lines. The tendency towards
those solutions is visible, reaching them exactly is however associated
with too high computational costs (see the discussion in section \ref{subsec:NumericalLimitations}).
The evolution of fracture opening and net pressure is plotted along
the centerline (e.g. $x=0$) in figures \ref{fig:BuoyantAfter}a)
and b). The head is identified once it departs from the source before
it subsequently shrinks. This shrinking makes the volume in the head
become negligible in comparison to the overall fluid volume after
sufficient buoyant propagation. When this moment is reached, the fracture
propagates in the viscosity-dominated regime (see also the nearly
self-similar footprint reported in the Supplementary Material). We
show in the supplementary material that the opening along the centerline
approaches the 2-D solution of \citet{RoLi05}. An approximated solution
may be possible when combining the zero toughness head (c.f. figure
7 of \citet{MoLe22c}) with the tail solution of \citet{RoLi07} (see
their equation 6.7), but is left for further study.

\subsection{Late time fracture shapes\label{subsec:Fracture shapes}}

\begin{figure}
\centering{}\includegraphics[width=1\textwidth]{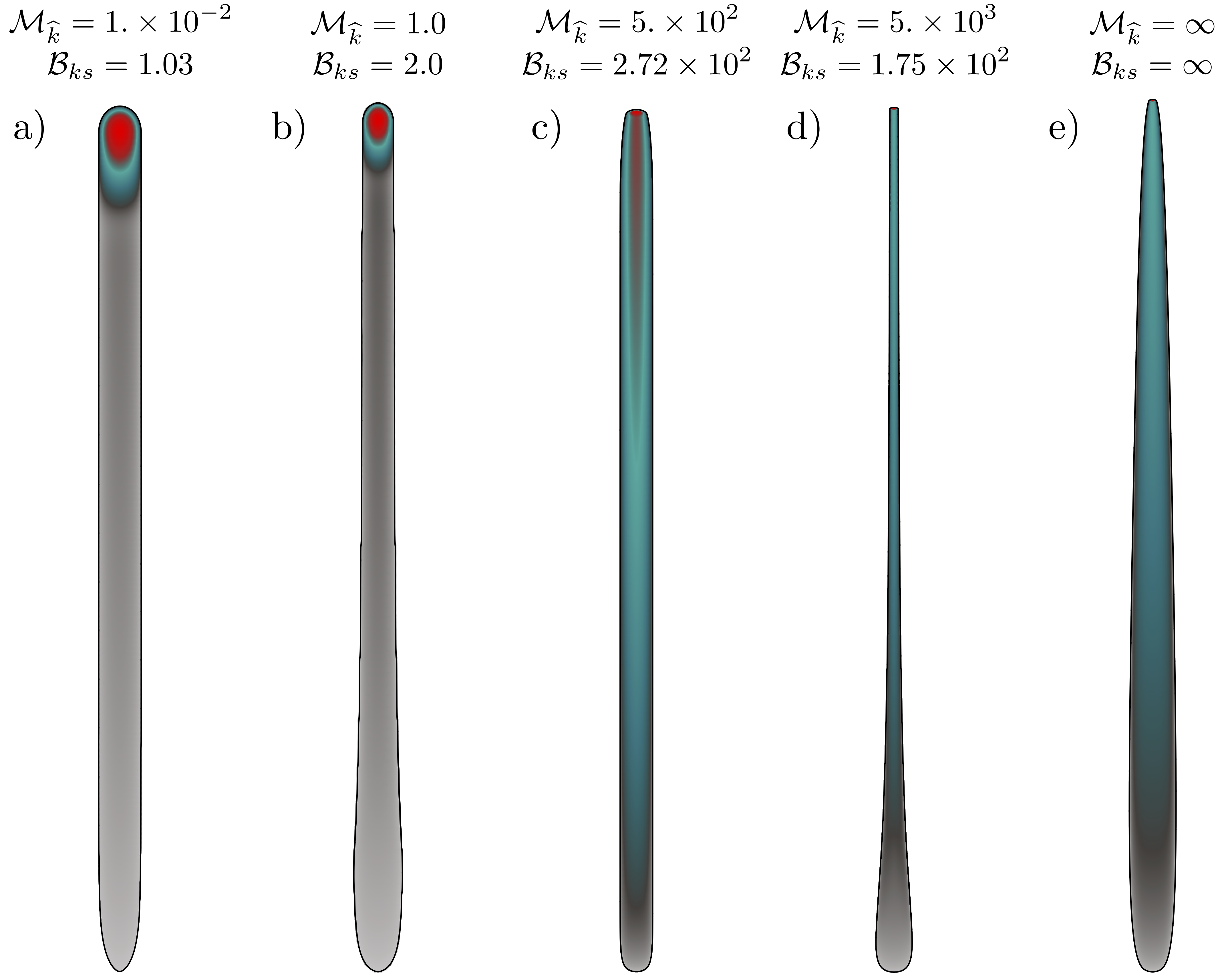}\caption{Phenotypes of possible buoyant hydraulic fractures of finite volume
emerging from a point source $\left(\mathcal{B}_{ks}\protect\geq1\right)$.
a) Toughness-dominated finger-like fracture (region 3 in figure \ref{fig:Sketch}).
b) Intermediate fracture with a stable breadth and negligible overrun.
c) Viscosity-dominated buoyant end of the release with stabilized
breadth (region 4 in figure \ref{fig:Sketch}). d) Viscosity-dominated
buoyant end of the release without stabilized breadth (region 5 in
figure \ref{fig:Sketch}). e) Zero-toughness case with a buoyant end
of the release $\left(\mathcal{B}_{ms}=10^{2}\right)$. a) and b)
are scaled by $\ell_{b}$ \citep{LiKe91}, c) to e) by $\ell_{m\widehat{m}}^{\left[V\right]}$
(see table \ref{Apptab:TransitionScales}).\label{fig:DifferentBuoyantShapes}}
\end{figure}
The governing mechanisms delimiting the different regions of the parametric
space of figure \ref{fig:ParametricSpace} give rise to different
phenotypes of fracture shape. Figure \ref{fig:DifferentBuoyantShapes}
displays the late-time shape of buoyant fractures in the different
regions (3-5) of the parametric space. Figure \ref{fig:DifferentBuoyantShapes}a)
shows the characteristic shape of a toughness-dominated buoyant fracture
at the end of the release (region 3). Their footprint is finger-like
with a constant breadth and head volume. Already early in the propagation,
the bulk of the released volume is located in the head (indicated
by the colour code). Except in the source region and for the expanding
head, no change in breadth is observed, and the overrun $O$ (see
equation (\ref{eq:Overrun})) is zero. For fractures with a uniform
breadth, not validating the toughness solution (e.g. $\mathcal{M}_{\widehat{k}}\in\left[10^{-2},10^{2}\right]$
and $\mathcal{B}_{ks}\geq1$, between regions 3 and 4), the bulk of
the fluid volume is similarly in the head. One difference is the change
in breadth observed close to the source region related to the end
of the release, giving rise to a small, non-zero overrun. When the
fractures are more viscosity-dominated (see figures \ref{fig:DifferentBuoyantShapes}c-e)),
the overrun becomes more pronounced, and the opening distribution
is more homogeneous along the fracture length. For example, figure
\ref{fig:DifferentBuoyantShapes}c) shows a viscosity-dominated, buoyant
fracture with a stabilized breadth at the end of the release (region
4) with a barely visible head (light red area at the propagating edge).
The red-coloured part extending long into the tail shows that the
tail opening is much closer to the head opening than in the toughness-dominated
cases of figures \ref{fig:DifferentBuoyantShapes}a) and b). The particularity
of this phenotype is its uniform breadth over a finite height due
to lateral stabilization (associated with a finite value of fracture
toughness). Figure \ref{fig:DifferentBuoyantShapes}d) (region 5)
emphasizes the approaching towards the late-time 3-D $K^{\left[V\right]}$
GG (2014) solution of viscosity-dominated fractures by the thinning
of the breadth along the fracture length towards its head. The head
breadth of this simulation still exceeds the limiting solution by
a factor of about $4.7$ and the opening distribution along the fracture
is still too homogeneous. In other words, a significant proportion
of the volume remains in the tail (compare the grey colour in figure
\ref{fig:DifferentBuoyantShapes}a) with the green colour in figure
\ref{fig:DifferentBuoyantShapes}d)). The last phenotypein figure
\ref{fig:DifferentBuoyantShapes}e), represents the case of a zero-toughness
simulation with a buoyant end of the release. Comparing this shape
to the zero toughness simulation with negligible buoyancy at the end
of the release (c.f. figures \ref{fig:BuoyantAfter}h-i)) reveals
no significant difference. All zero-toughness simulations, independent
of the state at the end of the release, will show this particular
shape. Only if a finite fracture toughness is present, the fracture
will tend to the late-time 3-D $K^{\left[V\right]}$ GG (2014) solution
and the shape will resemble figure \ref{fig:DifferentBuoyantShapes}d)
(see also figure 1b) of \citet{DaRi23}).

\section{Discussion}

\begin{figure}
\begin{centering}
\includegraphics[width=0.9\textwidth]{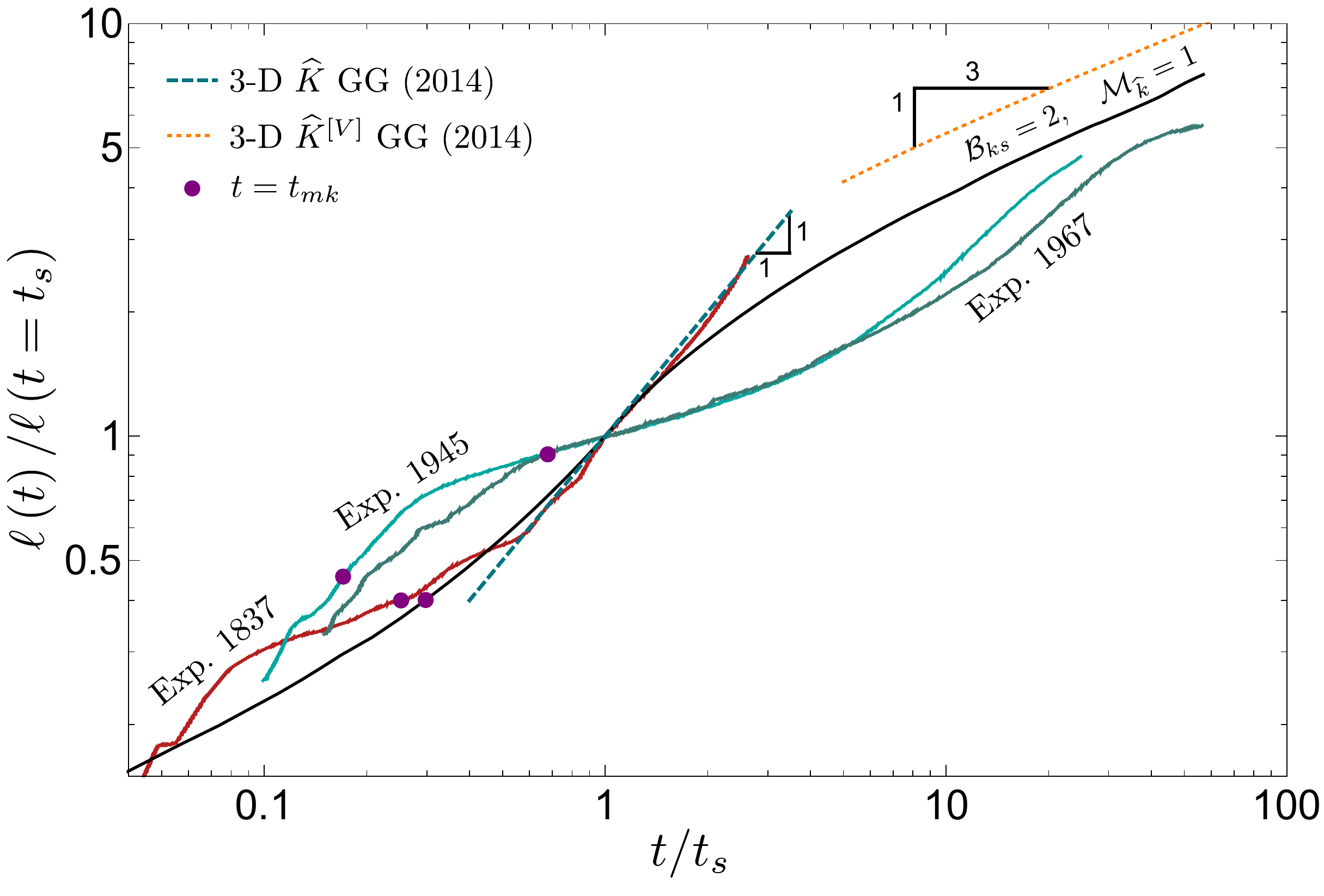}\caption{Fracture height evolution as a function of dimensionless time for
the experiments listed in table \ref{tab:ExperimentalData}. Data
extracted from figure 5a of \citet{DaRi23} based on experiments reported
in \citet{Smit19}. The black line shows a simulation with similar
dimensionless parameters to Exp. 1945 and Exp. 1967. Purple dots mark
the moment when the fracture becomes toughness dominated (e.g. $t=t_{mk}$),
and dashed lines indicate the limiting solutions derived by \citet{GaGe14}
respectively (green for a continuous release, orange for the release
of a finite volume).\label{fig:ExperimentComparison}}
\par\end{centering}
\end{figure}

\begin{table}
\begin{center}
\def~{\hphantom{0}}
\begin{tabular}{ccccc}

 & Unit & Exp. 1837 & Exp. 1945 & Exp. 1967\tabularnewline

$\mu_{f}$ & $\textrm{Pa}\cdot\textrm{s}$ & ${\displaystyle 1.74\times10^{-3}}$ & ${\displaystyle 48\times10^{-3}}$ & ${\displaystyle 970\times10^{-3}}$\\[3pt]

$E$ & $\textrm{Pa}$ & $1345$ & ${\displaystyle 426}$ & ${\displaystyle 944}$\\[2pt]

$\nu$ &  & $0.5$ & $0.5$ & $0.5$\\[2pt]

$K_{Ic}$ & $\textrm{Pa}\cdot\textrm{m}^{1/2}$ & ${\displaystyle 23.1}$ & $7.3$ & ${\displaystyle 16.2}$\\[2pt]

$\varDelta\rho$ & $\textrm{kg}\cdot\textrm{m}^{-3}$ & ${\displaystyle 260}$ & ${\displaystyle 160}$ & ${\displaystyle 150}$\\[2pt]

$V_{o}$ & $\textrm{m}^{3}$ & ${\displaystyle 2.\times10^{-5}}$ & ${\displaystyle 1.\times10^{-5}}$ & ${\displaystyle 1.\times10^{-5}}$\\[2pt]

$Q_{o}$ & $\textrm{m}^{3}\cdot\textrm{s}^{-1}$ & ${\displaystyle 1.23\times10^{-7}}$ & $8.33\times10^{-7}$ & ${\displaystyle 1.11\times10^{-7}}$\\[2pt]
 
$t_{s}$ & $\textrm{s}$ & $162$ & ${\displaystyle 12}$ & ${\displaystyle 90}$\\[2pt]

$\mathcal{M}_{\widehat{k}}$ &  & $1.20\times10^{-3}$ & $1.11$ & $0.76$\\[2pt]

$\mathcal{B}_{ks}$ &  & $2.28$ & $2.93$ & $1.24$\\[2pt]

$\mathcal{B}_{ms}$ &  & $57.7$ & $3.85$ & $1.49$\\[2pt]
\end{tabular}
\caption{Material parameters and the released volume $V_{o}$ are taken from
table 3 of \citet{DaRi23} (based on the work of \citet{Smit19}).
We extract the shut-in time from figure 5a of \citet{DaRi23} and
calculate the release rate as $Q_{o}=V_{o}/t_{s}$. \label{tab:ExperimentalData}}
\end{center}
\end{table}
We compare recent laboratory experiments with our scalings and numerical
simulations. We use three sets of parameters from experiments performed
by \citet{Smit19} and reported in \citet{DaRi23} (see table \ref{tab:ExperimentalData}).
The resulting dimensionless parameters are listed in table \ref{tab:ExperimentalData}.
It appears that these fractures are either toughness-dominated (experiment
1827) or in the transition with a uniform breadth (experiments 1945
and 1967). We report the evolution of fracture height with time in
figure \ref{fig:ExperimentComparison} (data of the experiments from
figure 5a of \citet{DaRi23}). Along with the three experiments, we
show our simulation closest to experiments 1945 and 1967 as well as
the limiting solutions derived by \citet{GaGe14}. The toughness-dominated
experiment (exp. 1837) display a linear fracture height growth with
time, expected from the continuous release scaling. Surprisingly,
the end of the release does not lead to a significant reduction in
height growth (c.f. the simulation with $\mathcal{M}_{\widehat{k}}=10^{-2}$
in figure \ref{fig:ToughnessHeadBreadthConvergence}b)), which continues
linearly until it reaches the top of the tank (end of the data stream).
We expect this to be related to free-surface effects attracting the
fracture, a hypothesis supported by observations of the other two
experiments. The fractures of the other experiments grow without showing
any scaling-based power laws. This behaviour is typical for many laboratory
experiments which unfortunately appear to be ``in-between'' limiting
regimes. Additionally, the extent of the hydraulic fractures created
often suffers from detrimental effects associated with the finite
size of the sample, making any comparison with theoretical and numerical
predictions difficult. The fact that the release rate in laboratory
experiments is often not constant, presents an additional inconvenience.
Especially in viscosity-dominated fracture propagation regimes, this
has a significant influence on fracture growth via $\mathcal{M}_{\widehat{k}}$.

The complete parametric space characterising 3-D finite volume buoyant
hydraulic fractures described in this paper should help in better
designing experiments within probing well-defined propagation history.

\section{Conclusions}

We have shown that finite-volume hydraulic fractures are entirely
characterized by a dimensionless buoyancy $\mathcal{B}_{ks}=\varDelta\gamma E^{\prime3/5}V_{o}^{3/5}/K_{Ic}^{8/5}$
relating the total released $V_{o}$ to the minimum volume necessary
for self-sustained buoyant propagation $V_{\widehat{k}}^{head}=K_{Ic}^{8/3}/\left(E^{\prime}\varDelta\gamma^{3/5}\right)$:
$\mathcal{B}_{ks}=\left(V_{o}/V_{\widehat{k}}^{\textrm{head}}\right)^{3/5}$
and a dimensionless viscosity $\mathcal{M}_{\widehat{k}}=\mu^{\prime}Q_{o}E^{\prime3}\varDelta\gamma^{2/3}/K_{Ic}^{14/3}$.
Although the emergence (or not) of a self-sustained buoyant fracture solely depends on $\mathcal{B}_{ks}$, in other
words, on the total volume released and material and fluid parameters, the details of the release (duration and injection rate) have a first order impact on the shape and propagation rates 
of the fracture through the dimensionless viscosity $\mathcal{M}_{\widehat{k}}$. 
Combining these  two dimensionless numbers $(\mathcal{B}_{ks},\,\mathcal{M}_{\widehat{k}})$, reveals six regions corresponding to distinct propagation
histories (see figure \ref{fig:ParametricSpace}).

For a finite value of the material fracture toughness ($K_{Ic}\neq0$), the toughness-dominated pulse solution of \citet{GaGe14}
characterizes the late-time buoyant head and the fracture breadth
in its vicinity ( $b^{head}=\pi^{-1/3}\ell_{b}$). Note that such a late-time solution may appear only at very late times.
In the zero-toughness   
case ($K_{Ic} = 0$), the fracture head shrinks eternally, and the maximum lateral breadth stabilizes at a finite value. It is thus possible to relate the limiting breadth
close to the head to the stabilized maximum one. We define this
parameter as the overrun $O$ and derive its value for the different
regions of the parametric space. Note that this parameter gives only
an idea of the shape:  a similar overrun does not imply that the
fracture has the same overall shape. When the fracturing toughness is zero, the head breadth tends to zero (e.g. $\ell_{b}=0\to b^{head}=0$), resulting in an infinite overrun. It is important to note that this does not imply unbounded lateral growth, as lateral growth is limited by the finite volume rather than fracture toughness. 

The identified late-time behaviour further fixes the late-time ascent
rate to the toughness-dominated solution as $\dot{\ell}_{\widehat{k}}^{\left[V\right]}\left(t\right)\sim V_{o}^{2/3}\varDelta\gamma^{7/9}/\left(K_{Ic}^{4/9}\mu^{\prime1/3}t^{2/3}\right)\propto t^{1/3}$
. An important observation is that the time power-law dependence of
the ascent rate for a viscosity-dominated buoyant fracture is equivalent
(e.g. $\dot{\ell}_{\widehat{m}}^{\left[V\right]}\sim V_{o}^{1/2}\varDelta\gamma^{1/2}/\left(E^{\prime1/6}\mu^{\prime1/3}t^{2/3}\right)\propto t^{1/3}$).
During its history, a buoyant hydraulic fracture can first ascent in a viscosity-dominated manner as
$\dot{\ell}_{\widehat{m}}^{\left[V\right]}\left(t\right)$ and then
transition to the limiting ascent rate dictated by the late toughness solution $\dot{\ell}_{\widehat{k}}^{\left[V\right]}\left(t\right)$.
The late-time ascent rate of the toughness limit is always faster (or at least equal) 
than the one of the viscosity-dominated limit ($\dot{\ell}{}_{\widehat{k}}^{\left[V\right]}\left(t\right)=\mathcal{B}_{ks}^{5/18}\dot{\ell}{}_{\widehat{m}}^{\left[V\right]}\left(t\right)$,
with $\mathcal{B}_{ks}\geq1$ for a self-sustained buoyant fracture).
Fractures transitioning when the fluid release is still ongoing can show
even higher velocities during their propagation history. Estimations
or averaging of vertical growth rates must be done with great care
and must necessarily account for both $\mathcal{M}_{\widehat{k}}$
and $\mathcal{B}_{ks}$. In other words, both the details of the release history (rate and duration) do significantly impact the ascent rate even long after the end of the release.
This implies that for realistic cases (as well as laboratory experiments), the detail of the released matter such that a more complicated evolution of the release (compared to the simple constant rate / finite duration) will certainly impact the growth of buoyant fractures.

It is noteworthy that most parameter combinations for natural or anthropogenic
hydraulic fractures would lead to self-sustained buoyant propagation
between the well-distinct regions of the parametric space depicted
in \ref{fig:ParametricSpace}. Additionally, the time required to
reach the late-time solution at the propagating edge and the fracture
size when doing so naturally clash with sample sizes in the laboratory
or the scales of heterogeneities in the upper lithosphere. We emphasize
that even though theoretically buoyant fractures emerge, nearly no
cases of buoyant fractures from hydraulic fracturing treatments reaching
the surface have been reported to our knowledge. We expect this contradiction to be related to the interaction
of the growing fracture with heterogeneities, fluid leak-off, and other possible arrest mechanisms
not accounted for in this contribution.  Accounting for these effects is
part of ongoing work and should elucidate why fractures ultimately arrest
at depth, even when $\mathcal{B}_{ks}$ is above unity.\\
\\

\noindent \textbf{Acknowledgements.} The authors gratefully acknowledge
in-depth discussions with Dmitry Garagash. \\
\textbf{Funding.} This work was funded by the Swiss National Science
Foundation under grant \#192237. \\
\textbf{Declaration of Interest. }The authors report no conflict of
interest.\\
\textbf{Data availability statement. }The data that support the findings
of this study are openly available at \href{https://doi.org/10.5281/zenodo.7788051}{10.5281/zenodo.7788051} .\\
\textbf{Author ORCID. }A. Möri, \href{https://orcid.org/0000-0002-7951-1238}{0000-0002-7951-1238};
B. Lecampion, \href{https://orcid.org/0000-0001-9201-6592}{0000-0001-9201-6592}\\
\textbf{Author contributions. }Andreas Möri: Conceptualization, Methodology,
Formal analysis, Investigation, Software, Validation, Visualization,
Writing - original draft. Brice Lecampion: Conceptualization, Methodology,
Formal analysis, Validation, Supervision, Funding acquisition, Writing
- review \& editing.

\newpage

\appendix

\section{Recapitulating tables of scales\label{Appsec:ScalingsTables}}

For completeness, we list all the characteristic scales used within
this contribution in the following tables. A Wolfram mathematica notebook
containing their derivation and the different scalings is also provided
as supplementary material.

\begin{table}
\begin{center}
\def~{\hphantom{0}}
\begin{tiny}
\begin{tabular}{ccccccc}
 & \multicolumn{2}{c}{\emph{\normalsize Radial}} & \multicolumn{4}{c}{\emph{\normalsize Elongated}}\\[3pt]

 & $\text{M}^{\left[V\right]}$ & $\text{K}^{\left[V\right]}$ & $\widehat{\text{M}}^{\left[V\right]}$ (tail) & $\widehat{\text{M}}^{\left[V\right]}$ (head) & $\widehat{\text{K}}^{\left[V\right]}$ (tail) & $\widehat{\text{K}}^{\left[V\right]}$ (head)\\[3pt]

$\ell_{*}^{\left[V\right]}$ & $\dfrac{E^{\prime1/9}V_{o}^{1/3}t^{1/9}}{\mu^{\prime1/9}}$ & $\dfrac{E^{\prime2/5}V_{o}^{2/5}}{K_{Ic}^{2/5}}$ & $\dfrac{V_{o}^{1/2}\varDelta\gamma^{1/2}t^{1/3}}{E^{\prime1/6}\mu^{\prime1/3}}$ & $\dfrac{E^{\prime11/24}V_{o}^{1/8}\mu^{\prime1/6}}{\varDelta\gamma^{5/8}t^{1/6}}$ & $\dfrac{V_{o}^{2/3}\varDelta\gamma^{7/9}t}{K_{Ic}^{4/9}\mu^{\prime1/3}}$ & $\ell_{b}$\\[2pt]

$b_{*}^{\left[V\right]}$ & $\ell_{*}^{\left[V\right]}$ & $\ell_{*}^{\left[V\right]}$ & $\dfrac{E^{\prime1/4}V_{o}^{1/4}}{\varDelta\gamma^{1/4}}$ & $\ell_{*}^{head,\left[V\right]}$ & $\ell_{b}=\dfrac{K_{Ic}^{2/3}}{\varDelta\gamma^{2/3}}$ & $\ell_{*}$\\[2pt]

$w_{*}^{\left[V\right]}$ & $\dfrac{V_{o}^{1/3}\mu^{\prime2/9}}{E^{\prime2/9}t^{2/9}}$ & $\dfrac{K_{Ic}^{4/5}V_{o}^{1/5}}{E^{\prime4/5}}$ & $\dfrac{V_{o}^{1/4}\mu^{\prime1/3}}{E^{\prime1/12}\varDelta\gamma^{1/4}t^{1/3}}$ & $w_{*}^{tail,\left[V\right]}$ & $\dfrac{V_{o}^{1/3}\mu^{\prime1/3}}{K_{Ic}^{2/9}\varDelta\gamma^{1/9}t^{1/3}}$ & $\dfrac{K_{Ic}^{4/3}}{E^{\prime}\varDelta\gamma^{1/3}}$\\[2pt]

$V_{*}^{\left[V\right]}$ & $V_{o}$ & $V_{o}$ & $V_{o}-V_{*}^{head,\left[V\right]}$ & $\dfrac{E^{\prime5/6}V_{o}^{1/2}\mu^{\prime2/3}}{\varDelta\gamma^{3/2}t^{2/3}}$ & $V_{o}-V_{*}^{head,\left[V\right]}$ & $\dfrac{K_{Ic}^{8/3}}{E^{\prime}\varDelta\gamma^{5/3}}$\\[2pt]

$p_{*}^{\left[V\right]}$ & $\dfrac{E^{\prime2/3}\mu^{\prime1/3}}{t^{1/3}}$ & $\dfrac{K_{Ic}^{6/5}}{E^{\prime1/5}V_{o}^{1/5}}$ & $\dfrac{E^{\prime2/3}\mu^{\prime1/3}}{t^{1/3}}$ & $\dfrac{E^{\prime11/24}V_{o}^{1/8}\mu^{\prime1/6}\varDelta\gamma^{3/8}}{t^{1/6}}$ & $\dfrac{E^{\prime}\varDelta\gamma^{5/9}V_{o}^{1/3}\mu^{\prime1/3}}{K_{Ic}^{8/9}t^{1/3}}$ & $K_{Ic}^{2/3}\varDelta\gamma^{1/3}$\\[2pt]

\multirow{2}{*}{$\mathcal{P}_{s}^{\left[V\right]}$} & $\mathcal{K}_{m}^{\left[V\right]}=(t/t_{mk}^{\left[V\right]})^{5/18}$ & $\mathcal{M}_{k}^{\left[V\right]}=(t/t_{mk}^{\left[V\right]})^{-1}$ & \multicolumn{2}{c}{$\mathcal{K}_{\widehat{m}}^{\left[V\right]}=\left(\mathcal{M}_{\widehat{k}}^{\left[V\right]}\right)^{-1/4}\mathcal{B}_{ks}^{5/48}$} & \multicolumn{2}{c}{$\mathcal{M}_{\widehat{k}}^{\left[V\right]}=\mu^{\prime}\dfrac{V_{o}E^{\prime3}\varDelta\gamma^{2/3}}{K_{Ic}^{14/3}t}$}\\[2pt]

 & $\mathcal{B}_{m}^{\left[V\right]}=(t/t_{m\widehat{m}}^{\left[V\right]})^{4/9}$ & $\mathcal{B}_{k}^{\left[V\right]}=\varDelta\gamma\dfrac{E^{\prime3/5}V_{o}^{3/5}}{K_{Ic}^{8/5}}$ & \multicolumn{2}{c}{} & \multicolumn{2}{c}{}\\
\end{tabular}
\end{tiny}
\caption{Characteristic scales (and governing dimensionless parameters) in
the different scalings. \label{Apptab:RadialScales}}
\end{center}
\end{table}

\begin{table}
\begin{center}
\def~{\hphantom{0}}
\begin{scriptsize}
\begin{tabular}{ccccc} 
 & $t^{\left[V\right]}$ & $\ell_{*}^{\left[V\right]}=b_{*}^{\left[V\right]}$ & $w_{*}^{\left[V\right]}$ & $p_{*}^{\left[V\right]}$\\[3pt]

$\text{M}^{\left[V\right]}\rightarrow\text{K}^{\left[V\right]}$ & $t_{mk}^{\left[V\right]}=\dfrac{E^{\prime13/5}V_{o}^{3/5}\mu^{\prime}}{K_{Ic}^{18/5}}$ & $\ell_{mk}^{\left[V\right]}=\dfrac{E^{\prime2/5}V_{o}^{2/5}}{K_{Ic}^{2/5}}$ & $w_{mk}^{\left[V\right]}=\dfrac{K_{Ic}^{4/5}V_{o}^{1/5}}{E^{\prime4/5}}$ & $p_{mk}^{\left[V\right]}=\dfrac{K_{Ic}^{6/5}}{E^{\prime1/5}V_{o}^{1/5}}$\\[2pt]

$\text{M}^{\left[V\right]}\rightarrow\widehat{\text{M}}^{\left[V\right]}$ & $t_{m\widehat{m}}^{\left[V\right]}=\dfrac{E^{\prime5/4}\mu^{\prime}}{V_{o}^{3/4}\varDelta\gamma^{9/4}}$ & $\ell_{m\widehat{m}}^{\left[V\right]}=\dfrac{E^{\prime1/4}V_{o}^{1/4}}{\varDelta\gamma^{1/4}}$ & $w_{m\widehat{m}}^{\left[V\right]}=\dfrac{V_{o}^{1/2}\varDelta\gamma^{1/2}}{E^{\prime1/2}}$ & $p_{m\widehat{m}}^{\left[V\right]}=E^{\prime1/4}V_{o}^{1/4}\varDelta\gamma^{3/4}$\\[2pt]

$\widehat{\text{M}}^{\left[V\right]}\rightarrow\text{\ensuremath{\widehat{\text{K}}}}^{\left[V\right]}$
(tail) & \multirow{2}{*}{$t_{\widehat{m}\widehat{k}}^{\left[V\right]}=\dfrac{E^{\prime11/4}V_{o}^{3/4}\varDelta\gamma^{1/4}\mu^{\prime}}{K_{Ic}^{4}}$} & $\ell_{\widehat{m}\widehat{k}}^{\left[V\right]}=\dfrac{E^{\prime1/4}V_{o}^{1/4}}{\varDelta\gamma^{1/4}}$ & $w_{\widehat{m}\widehat{k}}^{\left[V\right]}=\dfrac{K_{Ic}^{4/3}}{E^{\prime}\varDelta\gamma^{1/3}}$ & $p_{\widehat{m}\widehat{k}}^{\left[V\right]}=\frac{K_{Ic}^{4/3}}{E^{\prime1/4}V_{o}^{1/4}\varDelta\gamma^{1/12}}$\\[2pt]

$\widehat{\text{M}}^{\left[V\right]}\rightarrow\text{\ensuremath{\widehat{\text{K}}}}^{\left[V\right]}$
(head) &  & $\ell_{\widehat{m}\widehat{k}}^{head,\left[V\right]}=\ell_{b}=\dfrac{K_{Ic}^{2/3}}{\varDelta\gamma^{2/3}}$ & $w_{\widehat{m}\widehat{k}}^{head,\left[V\right]}=w_{\widehat{m}\widehat{k}}^{\left[V\right]}$ & $p_{\widehat{m}\widehat{k}}^{head,\left[V\right]}=K_{Ic}^{2/3}\varDelta\gamma^{1/3}$\\[2pt]

\end{tabular}
\end{scriptsize}
\caption{Transition scales between regimes. The transition scales of the $\text{M}^{\left[V\right]}\rightarrow\text{K}^{\left[V\right]}$
transition correspond to the $\text{K}^{\left[V\right]}$-scales,
and the transition scales of the $\widehat{\text{M}}^{\left[V\right]}\rightarrow\text{\ensuremath{\widehat{\text{K}}}}^{\left[V\right]}$
(head) to the $\widehat{\text{K}}^{\left[V\right]}$ scales of the
head, given respectively as the $\text{K}^{\left[V\right]}$ and $\widehat{\text{K}}^{\left[V\right]}$
(head) in table \ref{Apptab:RadialScales}.\label{Apptab:TransitionScales}}
\end{center}
\end{table}

\bibliographystyle{jfm}
\bibliography{ms.bbl}

\end{document}